\documentclass[epj,fleqn]{svjour}

\usepackage{latexsym}
\usepackage{graphics}
\usepackage{amsmath}
\usepackage{amsfonts}
\usepackage{amssymb}
\usepackage[dvips]{color}
\usepackage{url}
\usepackage{hhline}
\usepackage{subfigure}

\begin{document}
\title{Helicity amplitudes and electromagnetic decays of hyperon resonances}

\author{{Van Cauteren}
  Tim\inst{1,}\thanks{\email{Tim.VanCauteren@UGent.be}} \and Ryckebusch
Jan\inst{1} \and Metsch Bernard\inst{2} \and Petry Herbert-R.\inst{2}}

\institute{Ghent University, Dept. of Subatomic and Radiation Physics,
Proeftuinstraat 86, B-9000 Gent, Belgium \and Helmholtz Institut f\"ur
Strahlen- und Kernphysik, Nu{\ss}allee 14-16, D-53115 Bonn, Germany}

\date{\today}

\abstract{
We present results for the helicity amplitudes of the lowest-lying
hyperon resonances $Y^*$, computed within the framework of the Bonn
constituent-quark model, which is based on the Bethe-Salpeter
approach. The seven parameters entering the model were fitted to the
best known baryon masses. Accordingly, the results for the helicity
amplitudes are genuine predictions. Some hyperon resonances are seen
to couple more strongly to a virtual photon with finite $Q^2$ than to
a real photon. Other $Y^*$'s, such as the $S_{01}(1670)$ $\Lambda$
resonance or the $S_{11}(1620)$ $\Sigma$ resonance, couple very
strongly to real photons. We present a qualitative argument for
predicting the behaviour of the helicity asymmetries of baryon
resonances at high $Q^2$.
}

\maketitle

\def\Id{{\rm 1\kern-.3em I}}
\def\ud{\mathrm{d}}

\def\gpY{ $p(\gamma,K^+)Y$ }
\def\epY{ $p(e,e'K^+)Y$ }
\def\ggsY{ $p(\gamma^{(*)},K)Y$ }
\def\gsY{ $p(\gamma^*,K)Y$ }
\def\eL{ $p(e,e'K^+)\Lambda$ }
\def\ggsL{ $p(\gamma^{(*)},K)\Lambda$ }
\def\eY{ $p(e,e'K)Y$ }
\def\eS{ $p(e,e'K)\Sigma$ }

\def\Pbar{\overset{\rule{2mm}{.2mm}}{P}}
\def\Psibar{\overset{\rule{2mm}{.2mm}}{\Psi}}
\def\chibar{\overset{\rule{2mm}{.2mm}}{\chi}}
\def\Kbar{\overset{\rule{2mm}{.2mm}}{K}}
\def\qbar{\overset{\rule{2mm}{.2mm}}{q}}
\def\Gbar{\overset{\rule{2mm}{.2mm}}{\Gamma}}

\def\psl{p\hspace{-0.2cm}/}

\def\mM{\mathcal{M}}
\def\mW{\mathcal{W}}
\def\mD{\mathcal{D}}
\def\mF{\mathcal{F}}
\def\mC{\mathcal{C}}
\def\mD{\mathcal{D}}

\def\cD{\cal{D}}

\section{Introduction}\label{sec:intro_gen}

The present work is part of an effort to develop a consistent
description of kaon production processes of the type \gpY and
\epY~\cite{stijn12,stijn3,stijn4}. Recent data for these
processes are due to the CLAS Collaboration at Jefferson
Laboratory~\cite{CLAS03b}, the LEPS Collaboration at
SPring-8~\cite{zegers03}, and the SAPHIR Collaboration at
ELSA~\cite{saphir2003}. The abundant amount of new data calls for an
adequate theoretical treatment. The availability of such a model
appears indispensable for a proper interpretation of the experimental
results, spanning an energy range from threshold up to $~2.6$~GeV.

One of the major sources of theoretical uncertainties when modeling
\ggsY reactions, is the strength of the electromagnetic (EM) couplings
involved. This holds especially true for kaon electroproduction, where
the EM coupling depends on $Q^2$, the squared four-momentum transfered
by the virtual photon. The $Q^2$ dependence of the EM form factors is
largely unknown for the ``strange'' baryons~\cite{stijn4}.

\begin{figure}
\begin{center}
\resizebox{0.25\textwidth}{!}{\includegraphics{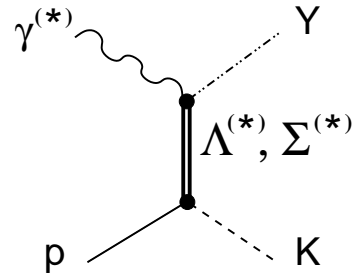}}
\caption{The $u$-channel diagram with an exchanged hyperon or hyperon
    resonance in a tree-level isobar model for kaon production. The
    photon couples to the intermediate $Y^{(*)}$ ($\Lambda^{(*)}$,
    $\Sigma^{(*)}$), resulting in the outgoing (ground-state) hyperon
    $Y$.}
\label{diag:u_chan_tree}
\end{center}
\end{figure}

In a tree-level description of kaon electroproduction, the $\gamma^* -
Y^{(*)}$ coupling comes into play in the $u$-channel (see
fig.~\ref{diag:u_chan_tree}). The electromagnetic vertex is
parametrized with the aid of elastic or transition form factors, which
are input to isobar models. In ref.~\cite{tim1}, we have presented our
results for the elastic form factors of ground-state hyperons and the
form factors of the $\Sigma^0(1193) \to \Lambda(1116)$ electromagnetic
transition, as computed in the framework of the Bonn Constituent-Quark
(CQ) model~\cite{loeringphd,loering1,merten1}. In this work, we focus
on the helicity amplitudes of hyperon resonances which decay
electromagnetically to the ground-state $\Lambda$ and $\Sigma$
hyperons. These amplitudes are calculated in a parameter-free manner,
and are compared with the (scarce) data to test the predictive power
of the Bonn CQ model.

The Bonn CQ model also provides the EM form factors of other
hadrons. Previous work has been reported for mesons and for nonstrange
baryons. For the pseudoscalar- and vector-meson elastic and transition
form factors~\cite{koll00}, a fair description was reached both for
data in the time-like~\cite{time_meson} and in the
space-like~\cite{space_meson} region.  For the pion, the outcome of
the calculations was reasonable, considering the high values for the
CQ masses in the model. For the nonstrange baryons~\cite{merten1}, the
results for the form factors and helicity amplitudes are reasonable to
excellent. Experimental data for the EM couplings of the nonstrange
baryons can be found in refs.~\cite{nucleon,jones_gayou} for the
nucleon, in refs.~\cite{delta} for the $\Delta$ resonance, and in
refs.~\cite{nonstr_res} for the lowest-lying nucleon and $\Delta$
resonances.

There have been several attempts to predict the EM properties of
hyperons and hyperon resonances. For the resonances, a number of
theoretical studies for the photo- and helicity amplitudes have been
performed since the beginning of the eighties. These include studies
of the EM decay widths and helicity amplitudes of the lowest-lying
hyperon resonances ($S_{01}(1405)$ and $D_{03}(1520)$) to the octet
($\Lambda(1116)$ and $\Sigma(1192)$) and decuplet ($\Sigma^*(1385)$)
ground states within the context of a nonrelativistic CQ model and a
bag model~\cite{darewych_kaxiras}. A treatment within the framework of
the chiral bag model was presented in the early nineties by Umino and
Myhrer in ref.~\cite{umino}. More recent approaches adopt
lattice QCD~\cite{leinweber93}, heavy-baryon chiral perturbation
theory~\cite{butler93}, the bound-state soliton
model~\cite{schat95a,schat95b}, the Skyrme
model~\cite{abada_haberichter}, and the chiral constituent quark
mo\-del~\cite{wagner}. Most of these model calculations are restricted
to the first and second hyperon resonance region (decuplet hyperons,
$S_{01}(1405)$, and $D_{03}(1520)$).  Note that the data on EM
couplings indeed only cover those states~\cite{PDG2004}. Results for
the other resonances are not constrained by data and should be
interpreted as predictions or extrapolations.

Since only static EM properties of the lowest-lying hyperon resonances
have been measured (the EM partial decay width), most of the
aforementioned studies did not consider the $Q^2$-dependence of the
helicity amplitudes. In addition, the validity of some models at
intermediate and high momentum transfers is rather questionable. For
$Q^2 \simeq m^2 \simeq {m^*}^2$, the hadron velocities in the lab
frame is of the order $v^2/c^2 \simeq 5/9$. This hints at sizeable
boost effects and at the necessity of a Lorentz-covariant model. Also
the validity of models based on chiral perturbation theory is
restricted to momenta transfers smaller than a certain parameter,
typically of the order of the mass of the nucleon.

In sect.~\ref{sec:BS_approach}, we will sketch how to compute baryon
properties within the framework of the Bonn CQ model. This model is
based on the Bethe-Salpeter approach, in which baryons are
characterized by their Bethe-Salpeter amplitudes. The equation obeyed
by this amplitude is presented in sect.~\ref{sec:BSE}. Solving this
equation is far from trivial. Yet for instantaneous interactions (as
used in the Bonn CQ model), the Bethe-Salpeter amplitude can be
derived from the Salpeter amplitude. As discussed in
sect.~\ref{sec:SE}, this equation is more easily solvable. The
calculation of the electromagnetic response of a hyperon resonance is
the topic of sect.~\ref{sec:response}. Section~\ref{sec:CME} is
devoted to the derivation of an expression for the current matrix
elements within the Bonn model. The current matrix elements are then
related to the helicity amplitudes of a hyperon resonance in
sect.~\ref{sec:HA}. In sect.~\ref{sec:lambda}, we will present the
helicity amplitudes of the lowest-lying $\Lambda$ resonances for $0
\leq Q^2 \leq 6$~GeV$^2$, decaying electromagnetically to the
$\Lambda(1116)$~(sect.~\ref{sec:lam*_lam}) and to the
$\Sigma(1193)$~(sect.~\ref{sec:lam*_sig}) octet hyperons. The helicity
amplitudes of the lowest-lying $\Sigma$ resonances
(sect.~\ref{sec:sigma}) are presented in sect.~\ref{sec:sig*_lam} for
the $\Sigma^* + \gamma^{(*)} \to \Lambda(1116)$ process and in
sect.~\ref{sec:sig*_sig} for the $\Sigma^* + \gamma^{(*)} \to
\Sigma(1193)$ process. In sect.~\ref{sec:hel_asym}, we will discuss
the computed helicity asymmetries, especially for large $Q^2$. In
sect.~\ref{sec:conclusions}, we present our conclusions. The effective
quark-quark interactions used in the Bonn model are given in
Appendix~\ref{sec:potentials}.

\begin{figure*}
\begin{center}
\resizebox{0.9\textwidth}{!}{\includegraphics{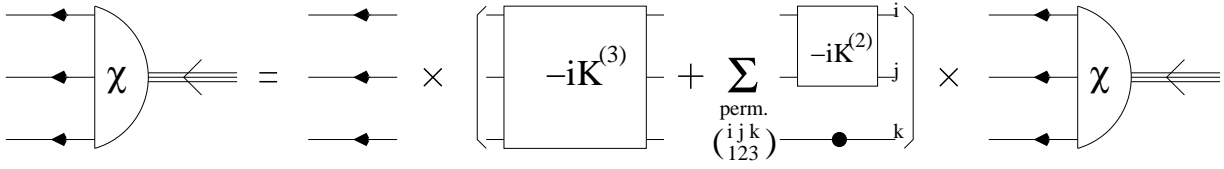}}
\caption{The BS equation in a schematic form. Arrows represent
quark propagators, a filled dot denotes an inverse propagator.}
\label{diag:BSE}
\end{center}
\end{figure*}

\section{Baryons in the Bonn model}\label{sec:BS_approach}

The Bethe-Salpeter (BS) formalism outlined here, is based on the
discussion of Le~Yaouanc \textit{et~al.}~\cite{yaouanc88}. It was
described in great detail and applied to mesons and baryons in
refs.~\cite{loering1,merten1,koll00}. In the Bonn CQ model, baryons
are considered to be composed of three CQ's. The three-CQ bound state
is described by the BS amplitude. The basic idea of the formalism is
to relate $n$-point Green's functions to the BS amplitudes of the
particles under consideration. Through an ingenious application of the
time-ordering operator, one isolates from the Green's function those
terms, which contain poles at those values of the kinematic variables
where the particles are on shell.  The residues of the Green's
function at the poles are the products of the bound state BS
amplitudes and their adjoints. It then boils down to finding an
equation for the Green's function which can be solved consistently to
a certain order in the coupling constants of the interactions. In the
course of this work, zeroth- and first-order approximations will be
adopted.

\subsection{The Bethe-Salpeter equation}\label{sec:BSE}

In the Bonn CQ model, the basic quantity describing a baryon
is the three-quark BS amplitude~:
\begin{multline}
\chi_{\Pbar,a_1,a_2,a_3} (x_1,x_2,x_3) \; \equiv \\ \langle 0 | T
\bigl( \Psi_{a_1}(x_1) \Psi_{a_2}(x_2) \Psi_{a_3}(x_3) \bigr) | \Pbar
\rangle \; ,
\label{eq:BSA}
\end{multline}
where $T$ is the time-ordering operator acting on the Heisenberg
quark-field operators $\Psi_{a_i}$, and $\Pbar$ is the total
four-mo\-men\-tum of the baryon with $\Pbar \cdot \Pbar \equiv
\Pbar_\mu \Pbar^\mu = M^2$.  The $a_i$ denote the quantum numbers in
Dirac, flavour and colour space. For the sake of conciseness, these
quantum numbers are frequently suppressed.

The BS amplitude is the solution of the BS equation for three
interacting relativistic
particles~\cite{loeringphd,loering1,salpeter}. In momentum space, this
equation reads~:
\begin{equation}
\chi_{\Pbar} \; = \; -i \, G^{(6)}_{0 \Pbar} \, \left( K^{(3)}_{\Pbar} +
\Kbar^{(2)}_{\Pbar} \right) \chi_{\Pbar} \; ,
\label{eq:BSE}
\end{equation}
where $\Pbar$ is the on-shell momentum. In the equation above the
arguments and indices have been suppressed. It is tacitly assumed that
one integrates over arguments and sums over indices that occur
twice. The diagrammatical analogue of the BS equation for the
amplitudes is shown in fig.~\ref{diag:BSE}. The normalization for the
BS amplitudes is given by~\cite{loeringphd,loering1}~:
\begin{equation}
-i \, \chibar_{\Pbar} \, \biggl[ \, P^\mu \, \frac{\partial} {\partial
 P^\mu} \left( {G^{(6)}_{0 P}}^{-1} + i K_P \right) \, \biggr]_{P =
 \Pbar} \, \chi_{\Pbar} \, = \, 2 M^2\; .
\label{eq:norm_cond_BS}
\end{equation}

In eq.~(\ref{eq:BSE}) $G^{(6)}_{0 \Pbar}$ is the direct product of the
dressed propagators of the three quarks~:
\begin{multline}
G^{(6)}_{0 \Pbar} (p_{\xi},p_{\eta};p'_{\xi},p'_{\eta}) \, = \, S^1_F
\left( \frac{1}{3} \Pbar + p_{\xi} + \frac{1}{2} p_{\eta} \right) \\
\otimes \, S^2_F \left( \frac{1}{3} \Pbar - p_{\xi} + \frac{1}{2}
p_{\eta} \right) \, \otimes \, S^3_F \left( \frac{1}{3} \Pbar -
p_{\eta} \right) \\ \times \, \left( 2 \pi \right)^4 \delta^{(4)}
\left( p_{\xi} - p'_{\xi} \right) \left( 2 \pi \right)^4 \delta^{(4)}
\left( p_{\eta} - p'_{\eta} \right) \; ,
\label{eq:G0P}
\end{multline}
where the arguments are Jacobi momenta, as defined in
ref.~\cite{tim1}.  The propagators $S^i_F$ (with $i=1,2,3$) are
approximated by those for constituent quarks~:
\begin{equation}
S^i_F(p_i) = \frac{i}{{\psl}_i-m_i+i\epsilon} \; ,
\label{eq:freeprop}
\end{equation}
where $m_i$ is the effective mass of the $i$'th constituent quark.

The quantity denoted by $K^{(3)}_{\Pbar}$ in eq.~(\ref{eq:BSE}) is the
three-particle irreducible interaction kernel for on-shell momenta
$\Pbar$. Further, $\Kbar^{(2)}_{\Pbar}$ is a sum of two-particle
irreducible interaction kernels, each multiplied by the inverse
propagator of the spectator quark as can be seen in
fig.~\ref{diag:BSE} and the expression~:
\begin{multline}
\Kbar^{(2)}_{\Pbar} \left( p_\xi,p_\eta;p'_\xi,p'_\eta \right) \, = \,
K^{(2)}_{ \left( \frac{2}{3} \Pbar + p_\eta \right)} \left(
p_\xi,p'_\xi \right) \\ \otimes \, \biggl[ S^{3}_F \left( \frac{1}{3}
\Pbar - p_\eta \right) \biggr]^{-1} \, \left( 2 \pi \right)^4
\delta^{(4)} \left( p_\eta - p'_\eta \right) \\ \textit{ +
cycl. perm. in quarks (123)} \; .
\label{eq:barK2}
\end{multline}

In any CQ model, there exists some freedom with respect to the
plausible types of interactions between the constituent quarks. We
will use the instantaneous approximation. In the center-of-mass frame,
the instantaneous approximation implies that the interaction kernels
$K^{(3)}_{\Pbar}$ and $K^{(2)}_{(p_i+p_j)}$ are independent of the
energy components of the Jacobi-momenta~:
\begin{subequations}
\label{eq:kernels}
\begin{alignat}{2}
&K^{(3)}_{\Pbar} \left(p_{\xi},p_{\eta};p'_{\xi},p'_{\eta}\right)
\bigg|_{\Pbar=(M,\vec{0})} \text{\hspace{-0.5cm}}&\equiv &V^{(3)}
\left(\vec{p}_{\xi},\vec{p}_{\eta};
\vec{p'}_{\xi},\vec{p'}_{\eta}\right) \label{eq:kernels3} \; ,\\
&K^{(2)}_{(\frac{2}{3}\Pbar+p_{\eta})} \left(p_{\xi},p'_{\xi}\right)
\bigg|_{\Pbar=(M,\vec{0})} &\equiv &V^{(2)}
\left(\vec{p}_{\xi},\vec{p}'_{\xi}\right) \label{eq:kernels2} \; .
\end{alignat}
\end{subequations}
We should mention here that whenever a quantity is to be evaluated in
the rest frame of the baryon, we will indicate this by the index $M$,
to make it clear that in this case $\Pbar = (M,\mathbf{0})$.

The potentials used in our calculations are those of model
$\mathcal{A}$ in ref.~\cite{loering2}, since they provided the best
results for the baryon spectrum. The three-particle interaction is
given by a \emph{confinement} potential $V^{(3)}_{\text{conf}}$ which
rises linearly with interquark distances with an appropriate Dirac
structure to avoid phenomenologically unwanted spin-orbit effects and
as a residual interaction the \emph{'t Hooft Instanton Induced
Interaction} $V^{(2)}_{\text{III}}$ which acts between flavour
antisymmetric quark pairs only~\cite{loeringphd}. The interaction
potentials are discussed in more detail in
Appendix~\ref{sec:potentials}.

The BS equation~(\ref{eq:BSE}) and the normalization condition of
eq.~(\ref{eq:norm_cond_BS}) for the BS amplitudes are Lorentz
covariant. The transformation properties of the quantities involved
are well-known, so that if one can find a solution for \textit{e.g.} the
BS amplitude in one Lorentz frame, it can be determined in an
arbitrary frame. We will exploit the relativistic covariance of the
model extensively by calculating quantities in the baryon's
center-of-mass frame and boosting these to the desired frame in order
to evaluate matrix elements.

\begin{figure*}
\begin{center}
\resizebox{0.9\textwidth}{!}{\includegraphics{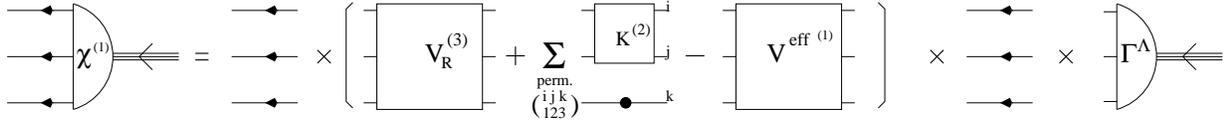}}
\caption{The reconstruction of the BS amplitude from the vertex
function according to eq.~(\ref{eq:approxBS}).}
\label{diag:BSE2}
\end{center}
\end{figure*}

\subsection{Reduction to the Salpeter equation}\label{sec:SE}

The problem of solving eq.~(\ref{eq:BSE}) is simplified by exploiting
the instantaneous property of the interaction kernels because the
integration over the energy components of the Jacobi momenta can be
performed analytically. This gives rise to a new object $\Phi_M$, the
Salpeter amplitude, which can be directly obtained from the full BS
amplitude~:
\begin{equation}
\Phi_M \left( \mathbf{p}_\xi,\mathbf{p}_\eta \right) \, = \, \int
\frac{\ud p^0_\xi}{\left( 2 \pi \right)} \frac{\ud p^0_\eta}{\left( 2
\pi \right)} \chi_M \left( (p^0_\xi,\mathbf{p}_\xi),
(p^0_\eta,\mathbf{p}_\eta) \right) \; .
\label{eq:SA}
\end{equation}
The integration over the energy components is easily performed in
situations where there are no genuine two-particle irreducible
interactions in eq.~(\ref{eq:BSE}), \textit{e.g.} for the ground-state
decuplet baryons which have symmetric spin wave functions. For other
ba\-ry\-ons, where the 't~Hooft instanton induced interaction
$V^{(2)}_{\text{III}}$ is non-vanishing, the inverse quark propagator
in the two-particle kernel (eq.~(\ref{eq:barK2})) introduces an extra
dependence on the energy components of the Jacobi momenta in the
right-hand side of eq.~(\ref{eq:BSE}). This makes it impossible to do
the integration analytically and a slightly different approach is
needed, as is explained in ref.~\cite{loering1} and in the Appendix of
ref.~\cite{merten1}. There, it is pointed out that for reconstructing
the Bethe-Salpeter amplitude defined in eq.~(\ref{eq:BSA}), it
suffices to compute the projection of the Salpeter amplitude of
eq.~(\ref{eq:SA}) onto the purely positive-energy and negative-energy
states. This can be accomplished in the standard manner by introducing
the energy-projection operators~:
\begin{equation}
\Lambda^{\pm}_i \left( \mathbf{p}_i \right) \, = \, \frac{ \omega_i \left(
  \mathbf{p}_i \right) \Id \pm H_i\left( \mathbf{p}_i \right) } { 2 \omega_i
  \left( \mathbf{p}_i \right)} \; ,
\label{eq:energyprojector}
\end{equation}
where $\omega _i (\mathbf{p}_i) = \sqrt{m_i^2 + |\mathbf{p}_i|^2}$ denotes
the energy and
\begin{equation}
H_i( \mathbf{p}_i ) \, = \, \gamma^0 \, ( \mathbf{\gamma} \cdot
\mathbf{p}_i + m_i ) \; ,
\end{equation}
is the free Dirac Hamiltonian for the $i$'th CQ.  With the above
definitions, one can project the Salpeter amplitude onto its purely
positive- and purely negative-energy components
\begin{multline}
\Phi^{\Lambda}_M \left( \mathbf{p}_\xi,\mathbf{p}_\eta \right) \, = \,
\left[ \Lambda^{+++} \left( \mathbf{p}_\xi,\mathbf{p}_\eta \right) +
\Lambda^{---} \left( \mathbf{p}_\xi,\mathbf{p}_\eta \right) \right] \\
\times \; \int \frac{\ud p^0_\xi} {\left( 2 \pi \right)} \frac{\ud
p^0_\eta} {\left( 2 \pi \right)} \chi_M \left(
(p^0_\xi,\mathbf{p}_\xi), (p^0_\eta,\mathbf{p}_\eta) \right) \; ,
\label{eq:projSA}
\end{multline}
where $\Lambda^{+++} \left( \mathbf{p}_\xi,\mathbf{p}_\eta \right) =
\Lambda^+_1 \left( \mathbf{p}_1 \right) \otimes \Lambda^+_2 \left(
\mathbf{p}_2 \right) \otimes \Lambda^+_3 \left( \mathbf{p}_3 \right)$
and $\Lambda^{---} \left( \mathbf{p}_\xi,\mathbf{p}_\eta \right) =
\Lambda^-_1 \left( \mathbf{p}_1 \right) \otimes \Lambda^-_2 \left(
\mathbf{p}_2 \right) \otimes \Lambda^-_3 \left( \mathbf{p}_3
\right)$. After a tedious calculation~\cite{loeringphd}, one obtains
an equation for the projected Salpeter amplitude, which is given by~:
\begin{multline}
\Phi^{\Lambda}_M \left( \mathbf{p}_\xi, \mathbf{p}_\eta \right) =
\biggl[ \frac{ \Lambda^{+++} \left( \mathbf{p}_\xi, \mathbf{p}_\eta
\right)} {M - \Omega \left( \mathbf{p}_\xi, \mathbf{p}_\eta \right) +
i \varepsilon} \phantom{\biggr]} \\ \phantom{\biggl[} + \, \frac{
\Lambda^{---} \left( \mathbf{p}_\xi, \mathbf{p}_\eta \right)} {M +
\Omega \left( \mathbf{p}_\xi, \mathbf{p}_\eta \right) - i \varepsilon}
\biggr] \; \gamma^0 \otimes \gamma^0 \otimes \gamma^0 \\
\times \, \int \frac{\ud^3 p'_\xi} {\left( 2 \pi \right)^3}
\frac{\ud^3 p'_\eta} {\left( 2 \pi \right)^3} \; V^{(3)} \left(
\mathbf{p}_\xi, \mathbf{p}_\eta; \mathbf{p}'_\xi, \mathbf{p}'_\eta
\right) \, \Phi^{\Lambda}_M \left( \mathbf{p}'_\xi, \mathbf{p}'_\eta
\right) \\ + \, \left[ \frac{ \Lambda^{+++} \left(
\mathbf{p}_\xi, \mathbf{p}_\eta \right)} {M - \Omega \left(
\mathbf{p}_\xi, \mathbf{p}_\eta \right) + i \varepsilon} \, - \,
\frac{\Lambda^{---} \left( \mathbf{p}_\xi, \mathbf{p}_\eta \right)} {M
+ \Omega \left( \mathbf{p}_\xi, \mathbf{p}_\eta \right) - i
\varepsilon} \right] \\ \times \, \int \frac{\ud^3 p'_\xi}
{\left( 2 \pi \right)^3} \; \biggl[ \left[ \, \gamma^0 \otimes
\gamma^0 \, V^{(2)} \left( \mathbf{p}_\xi, \mathbf{p}'_\xi \right)
\right] \otimes \Id \; \biggr] \, \Phi^{\Lambda}_M \left(
\mathbf{p}'_\xi, \mathbf{p}_\eta \right) \\ + \textit{
cycl. perm. in quarks (123)} \; ,
\label{eq:salpeq}
\end{multline}
where $\Omega \left( \mathbf{p}_\xi, \mathbf{p}_\eta \right)$ is the
sum of the energies of the three CQ's in the center-of-mass frame
\begin{equation}
\Omega \, = \, \sum ^{3}_{i = 1} \omega_i \, = \, \sum ^{3}_{i = 1} \sqrt{
  |\mathbf{p}_i|^2 + m^2_i} \; .
\label{eq:kin_energy}
\end{equation}
In principle, one would need the full Salpeter amplitude to
reconstruct the BS amplitude, but it turns out that the terms with the
smallest denominators are exactly those with projector structures
$\Lambda^{+++} \left( \mathbf{p}_\xi,\mathbf{p}_\eta \right)$ and
$\Lambda^{---} \left( \mathbf{p}_\xi,\mathbf{p}_\eta \right)$. The
denominators of the terms with other projector structures are large
enough, so that these terms may safely be neglected~\cite{loeringphd}.

Once the Salpeter equation~(\ref{eq:salpeq}) has been solved, the vertex
function $\Gamma^\Lambda_M$ can be constructed~:
\begin{multline}
\Gamma^\Lambda_M \left( \mathbf{p}_\xi, \mathbf{p}_\eta \right) \, =
\\ -i \, \int \frac{\ud^3 p'_\xi} {\left( 2 \pi \right)^3} \frac{\ud^3
p'_\eta} {\left( 2 \pi \right)^3} \, \biggl[ \, V^{(3)}_\Lambda \left(
\mathbf{p}_\xi, \mathbf{p}_\eta; \mathbf{p}'_\xi, \mathbf{p}'_\eta
\right) \phantom{\biggr]} \\ \phantom{\biggl[} + \, V^{eff^{(1)}}_M
\left( \mathbf{p}_\xi, \mathbf{p}_\eta; \mathbf{p}'_\xi,
\mathbf{p}'_\eta \right) \, \biggr] \, \Phi^{\Lambda,(1)}_M
\left( \mathbf{p}'_\xi, \mathbf{p}'_\eta \right) \; .
\label{eq:vertexfunc}
\end{multline}
At first order, this vertex function describes how the three CQ's
couple to form a baryon. It can be related to the BS amplitude
through~:
\begin{equation}
\chi_{\Pbar} \approx \chi^{(1)}_{\Pbar} = \left[ G_{0 \Pbar} \left( V^{(3)}_R +
\bar{K}^{(2)}_{\Pbar} - V^{eff^{(1)}}_{\Pbar} \right) G_{0
\Pbar} \right] \Gamma^\Lambda_{\Pbar} \; ,
\label{eq:approxBS}
\end{equation}
of which a diagram is shown in fig.~\ref{diag:BSE2}.

In eqs.~(\ref{eq:vertexfunc}) and~(\ref{eq:approxBS}), we have defined
\begin{multline}
V^{(3)}_\Lambda = \left( \gamma^0 \otimes \gamma^0 \otimes \gamma^0
\right) \ \left( \Lambda^{+++} \ + \ \Lambda^{---} \right) \\ \times
\, \left( \gamma^0 \otimes \gamma^0 \otimes \gamma^0 \right) \;
V^{(3)} \; \left( \Lambda^{+++} \ + \ \Lambda^{---} \right) \; ,
\label{eq:def_V3L}
\end{multline}
which is that part of the three-body potential which involves purely
positive-energy and negative-energy components of the amplitudes only.
Further, $V^{(3)}_R = V^{(3)} - V^{(3)}_\Lambda$ is the remaining part
which involves the mixed-energy components
only. $V^{eff^{(1)}}_{\Pbar}$ is a first-order approximation of an
effective potential with three-body structure which parameterizes the
two-body interaction~\cite{loering1,merten1}. Further,
$\Kbar^{(2)}_{\Pbar}$ is defined in eqs.~(\ref{eq:barK2})
and~(\ref{eq:kernels2}).

\begin{figure*}
\begin{center}
\resizebox{0.85\textwidth}{!}{\includegraphics{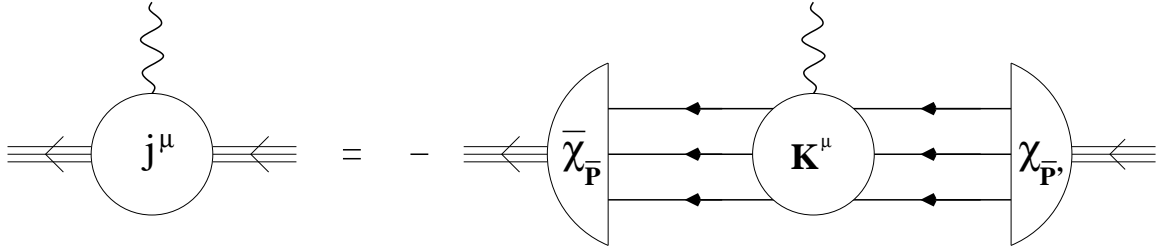}}
\caption{Diagrammatic representation of eq~(\ref{eq:CME_BSA})
for the CME.}
\label{diag:CME_BSA}
\end{center}
\end{figure*}

\section{Electromagnetic response}\label{sec:response}

In the Bonn CQ model, it is possible to calculate the matrix elements
of any operator which can be written in terms of quark-field creation
and annihilation operators. We will focus on the \emph{Current Matrix
Elements} (CME's) with one incoming and one outgoing baryon. The
current operator used in this work describes EM transitions. The
incoming and outgoing states, are bound states of three constituent
quarks, which are described by the BS amplitudes discussed in
sect.~\ref{sec:BS_approach}. In sect.~\ref{sec:CME}, we will derive
how the CME's are related to the BS amplitudes. In a next step, we
will derive the CME's in terms of the lowest-order approximations to
the BS amplitudes and the most simple expression for the photon
coupling to a three-quark bound state. In sect.~\ref{sec:HA}, the
computed current matrix elements will be linked to the helicity
amplitudes.

\subsection{Current matrix elements}\label{sec:CME}

After determining the BS amplitudes and the corresponding vertex
functions according to eqs.~(\ref{eq:vertexfunc})
and~(\ref{eq:approxBS}), the CME's can be computed through the
following definition of the current operator $j^\mu(x)$
\begin{equation}
\langle \, \Pbar \, | \, j^\mu(x) \, | \, \Pbar' \, \rangle \, = \,
\langle \, \Pbar \, | \, \Psibar (x) \, \hat{q} \, \gamma^\mu \, \Psi
(x) \, | \, \Pbar' \, \rangle \; ,
\label{eq:cur_mat_el1}
\end{equation}
where $\Psi$ and $\hat{q}$ are the CQ field in the Heisenberg picture
and the charge operator respectively. The current operator $j^\mu(x)$
corresponds to the photon coupling to a point-like CQ.  The above
matrix element can be expressed in terms of the objects defined in
sects.~\ref{sec:BSE} and~\ref{sec:SE}~\cite{merten1,mertenphd}~:
\begin{multline}
\langle \, \Pbar \, | \, j^\mu (0) \, | \, \Pbar' \, \rangle \; = \; -
\int \frac{\ud^4 p_\xi} {\left( 2 \pi \right)^4} \frac{\ud^4 p_\eta}
{\left( 2 \pi \right)^4} \frac{\ud^4 p'_\xi} {\left( 2 \pi \right)^4}
\frac{\ud^4 p'_\eta} {\left( 2 \pi \right)^4} \\ \times \,
\chibar_{\Pbar} \left( p_\xi,p_\eta \right) \; K^\mu_{P;q;P'} \left(
p_\xi,p_\eta;p'_\xi,p'_\eta \right) \, \chi_{\Pbar'} \left(
p'_\xi,p'_\eta \right) \; .
\label{eq:CME_BSA}
\end{multline}
This equation can be most easily interpreted with the aid of the
diagrams shown in fig.~\ref{diag:CME_BSA}.

\begin{figure*}
\begin{center}
\resizebox{0.85\textwidth}{!}{\includegraphics{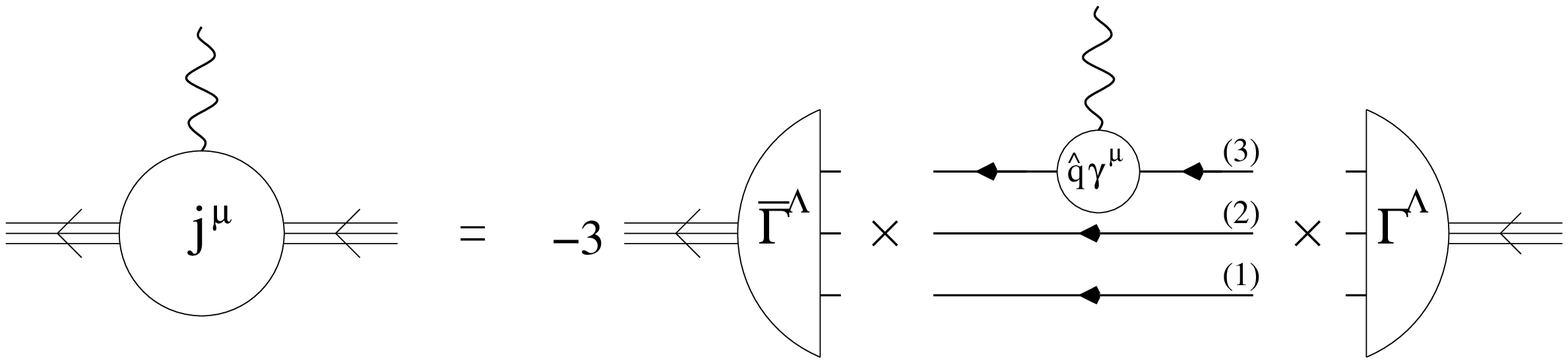}}
\caption{Feynman diagram showing the coupling of the photon to
the third CQ as in eq.~(\ref{eq:cur_mat_el2}). The other two CQ's are
spectators.}
\label{diag:current}
\end{center}
\end{figure*}

Up to this point, we have not introduced any approximation concerning
the order of the interactions. In acquiring the BS amplitudes within
the framework of sect.~\ref{sec:BS_approach}, however, we restricted
ourselves to the lowest order for the three-particle and two-particle
irreducible kernels in order to make the equations analytically
tractable and numerically computable. For a consistent calculation, a
lowest-order approximation for the kernel $K^\mu$ appears
necessary. The kernel $K^\mu$ in eq.~(\ref{eq:CME_BSA}) should thus be
expressed in terms of one-quark propagators and interaction kernels up
to lowest order. Using Wick's theorem, one can subsequently write all
connected terms without any interactions contributing to the
seven-point Green's function. We find 18 terms, which can be
subdivided into three groups, which are interconnected by a
permutation operator on the CQ's. However, the BS amplitudes are
antisymmetric by construction when interchanging two CQ's, so the
antisymmetric contributions in the kernel $K^\mu$ will be projected
out. Therefore, only three terms will have to be taken into account,
each of which describes the photon coupling to one of the CQ's. The
expression for $K^\mu$ up to zeroth order in configuration space
reads~\cite{merten1}~:
\begin{multline}
K^{\mu (0)} \left( x''_1,x''_2,x''_3;x;x'''_1,x'''_2,x'''_3 \right) \;
= \\ {S^1_F}^{-1} \left( x''_1 - x'''_1 \right) \, \otimes \,
{S^2_F}^{-1} \left( x''_2 - x'''_2 \right) \\ \otimes \, \left[
\delta^{(4)} \left( x''_3 - x \right) \hat{q} \gamma^\mu \,
\delta^{(4)} \left( x - x'''_3 \right) \right] \\ +
\textit{cycl. perm. in quarks (123)} \; . \label{eq:K_mu_0_conf}
\end{multline}
>From this, the expression for the kernel in momentum-space can be
easily obtained.

In evaluating eq.~(\ref{eq:CME_BSA}), the kernel $K^\mu_{P;q;P'}$ is
considered into lowest order, as in
eq.~(\ref{eq:K_mu_0_conf}). Exploiting the cyclic permutation
symmetry of the latter, the CME is obtained easily~:
\begin{multline}
\langle \, \Pbar \, | \, j^\mu (0) \, | \, \Pbar' \, \rangle \; \simeq
\; (-3) \int \frac{\ud^4 p_\xi} {\left( 2 \pi \right)^4} \frac{\ud^4
p_\eta} {\left( 2 \pi \right)^4} \; \chibar^{(1)}_{\Pbar} \left(
p_\xi,p_\eta \right) \\ \times \biggl[ {S^1_F}^{-1} \left( \frac{P}{3}
+ p_\xi + \frac{p_\eta}{2} \right) \otimes {S^2_F}^{-1} \left(
\frac{P}{3} - p_\xi + \frac{p_\eta}{2} \right) \phantom{\biggr]} \\
\phantom{\biggl[} \otimes \, \hat{q} \gamma^\mu \biggr] \
\chi^{(1)}_{\Pbar'} \left( p_\xi,p_\eta + \frac{2}{3} q \right) \ ,
\label{eq:CME_0_BSA_K_BSA}
\end{multline}
where $q$ is the (incoming) photon four-momentum. In the above
equation, we are using the first order approximation to the BS
amplitudes from eq.~(\ref{eq:approxBS}). Instead of explicitly
calculating the BS amplitudes with eq.~(\ref{eq:approxBS}) and
inserting them into eq.~(\ref{eq:CME_0_BSA_K_BSA}), it is more
convenient to express the CME's in terms of the vertex functions. We
insert the vertex functions $\Gamma^\Lambda_{\Pbar}$ from
eqs.~(\ref{eq:vertexfunc}) and~(\ref{eq:approxBS}) into our
approximate formula for the CME and retain the lowest-order
terms. Eventually, we arrive at~\cite{merten1,mertenphd}~:
\begin{multline}
<\Pbar | j^\mu(0) | M> \; \simeq \; (-3) \int \frac{d^4p_\xi}
{(2\pi)^4} \frac{d^4p_\eta} {(2\pi)^4} \; \Gbar^\Lambda_{\Pbar} \left(
p_\xi, p_\eta \right) \\ \times \, S^1_F \left( \frac{M}{3} + p_\xi +
\frac{p_\eta}{2} \right) \, \otimes \, S^2_F \left( \frac{M}{3} -
p_\xi + \frac{p_\eta}{2} \right) \\ \otimes \, \left[ S^3_F \left(
\frac{M}{3} - p_\eta + q \right) \, \hat{q} \gamma^\mu \, S^3_F \left(
\frac{M}{3} - p_\eta \right) \right] \\ \times \, \Gamma^\Lambda_M
\left( p_\xi, p_\eta + \frac{2}{3} q \right) \; ,
\label{eq:cur_mat_el2}
\end{multline}
Here, $\Gbar^\Lambda_{\Pbar}$ is the adjoint vertex function,
calculated in the c.o.m. frame according to
\begin{equation}
\Gbar^\Lambda_{M} = - \left( \Gamma^\Lambda_{M}
\right)^\dagger \gamma^0 \otimes \gamma^0 \otimes \gamma^0 \; .
\label{eq:adj_vert_func}
\end{equation}
Under a Lorentz boost, this vertex function transforms
as~\cite{mertenphd}
\begin{equation}
\Gamma_{\Pbar} \left(p_\xi,p_\eta \right) \; = \; S^1_\Lambda \otimes
S^2_\Lambda \otimes S^3_\Lambda \; \Gamma_{\Lambda^{-1} \Pbar} \Big(
\Lambda^{-1} \bigl( p_\xi \bigr),\Lambda^{-1} \big( p_\eta \big) \Big)
\; , \label{eq:vert_func_boost}
\end{equation}
with $\Lambda$ the boost matrix and $S^i_\Lambda$ the corresponding
boost operator acting on the $i$'th quark (not to be confused with the
propagator of the $i$'th quark $S^i_F$).
Equation~(\ref{eq:cur_mat_el2}) is a consistent lowest-order
approximation of the CME. We refer the reader to refs.~\cite{merten1}
and~\cite{mertenphd} for more details, and to fig.~\ref{diag:current}
for a schematic representation of eq.~(\ref{eq:cur_mat_el2}).  After
an appropriate treatment of the pole terms in the integration over the
energy variables, in the remaining integral over $\mathbf{p}_\xi$ and
$\mathbf{p}_\eta $, the azimuthal dependence can be reduced to
$(\phi_\xi - \phi_\eta)$, leaving one with five-dimensional integrals,
which are computed numerically.

\subsection{Helicity amplitudes}\label{sec:HA}

The literature on EM decays of nonstrange baryon resonances within the
framework of a quark model is
extensive~\cite{merten1,capstick_close}. For resonances, the concept
of EM form factors as coefficients to EM-vertex structures is
involved, especially for spin $J \geq 3/2$ resonances (see \textit{e.g.}
ref.~\cite{devenish76} for $J=3/2$ resonances). In general, the EM
properties are parameterized in terms of \emph{helicity amplitudes
(HA's)}. These quantities can be directly written in terms of the
CME's of the constituent quark model.

Depending on the conventions made with respect to normalization
factors, other definitions for the HA's emerge. Using the conventions
of ref.~\cite{merten1}, one gets for the EM transitions between
excited ($B^*$) and ground-state ($B$) baryons~:
\begin{subequations}
\begin{multline}
A_{1/2} \left( B^* \rightarrow B \right) = \\ 
\mD \, 
\left\langle B, \Pbar, \frac{1}{2} \right| 
j^1 (0) + i \, j^2 (0) 
\left| B^*, \Pbar^*, -\frac{1}{2} \right\rangle \; , 
\label{eq:HA_def_a12}
\end{multline}
\begin{multline}
A_{3/2} \left( B^* \rightarrow B \right) = \\ 
\mD \, 
\left\langle B, \Pbar, -\frac{1}{2} \right| j^1 (0) + i \, j^2 (0) 
\left| B^*, \Pbar^*, -\frac{3}{2} \right\rangle \; , 
\label{eq:HA_def_a32}
\end{multline}
\begin{equation}
C_{1/2} \left( B^* \rightarrow B \right) \, = 
\, \mD \, 
\left\langle B, \Pbar, \frac{1}{2} \right| 
j^0 (0) 
\left| B^*, \Pbar^*, \frac{1}{2} \right\rangle \; ,
\label{eq:HA_def_c12}
\end{equation}
\label{eq:HA_def}
\end{subequations}
with $\mD = \sqrt{\frac {\pi \alpha} {2 m ({m^*}^2 - m^2)}}$. There
are only two independent CME's for $B^*(J^*=1/2) \rightarrow B(J=1/2)$
transitions, and three for $B^*(J^*\geq3/2) \rightarrow B(J=1/2)$
transitions. With the above normalizations, the width corresponding to
the EM decay of an excited state $B^*$ with mass $m^*$ to a ground
state baryon $B$ with mass $m$ and spin $J=1/2$, is given by~:
\begin{equation}
\Gamma_\gamma \ = \ \frac {|\mathbf{q}|^2} {4 \pi^2 \alpha} \ \frac {2m}
      {\left( 2 J^* + 1 \right) m^*} \ \left[ |A_{1/2}|^2 +
      |A_{3/2}|^2 \right] \; .
\label{eq:EM_decaywidth}
\end{equation}
Here, $|\mathbf{q}| = \frac{{m^*}^2-m^2}{2m^*}$ is the three-momentum
of the photon in the rest frame of the initial baryon resonance, and
$\alpha = \frac{e^2}{4\pi} \simeq \frac{1}{137}$ is the EM
fine-structure constant. The adopted definition for the EM decay width
differs from the one given by the \emph{Particle Data Group
(PDG)}~\cite{PDG2004} by a factor of $e^2 = 4 \pi \alpha$. The PDG
tables present the experimentally known EM decay widths and
photo-amplitudes $A_i \left( Q^2 \rightarrow 0 \right)$. We compute
the full $Q^2$ dependence of the EM properties of hyperon resonances
in terms of HA's in sects.~\ref{sec:lambda} and~\ref{sec:sigma}.

\section{Results for the $\Lambda$-resonances}\label{sec:lambda}

\begin{figure*}
\begin{center}
\resizebox{0.95\textwidth}{!}{\includegraphics{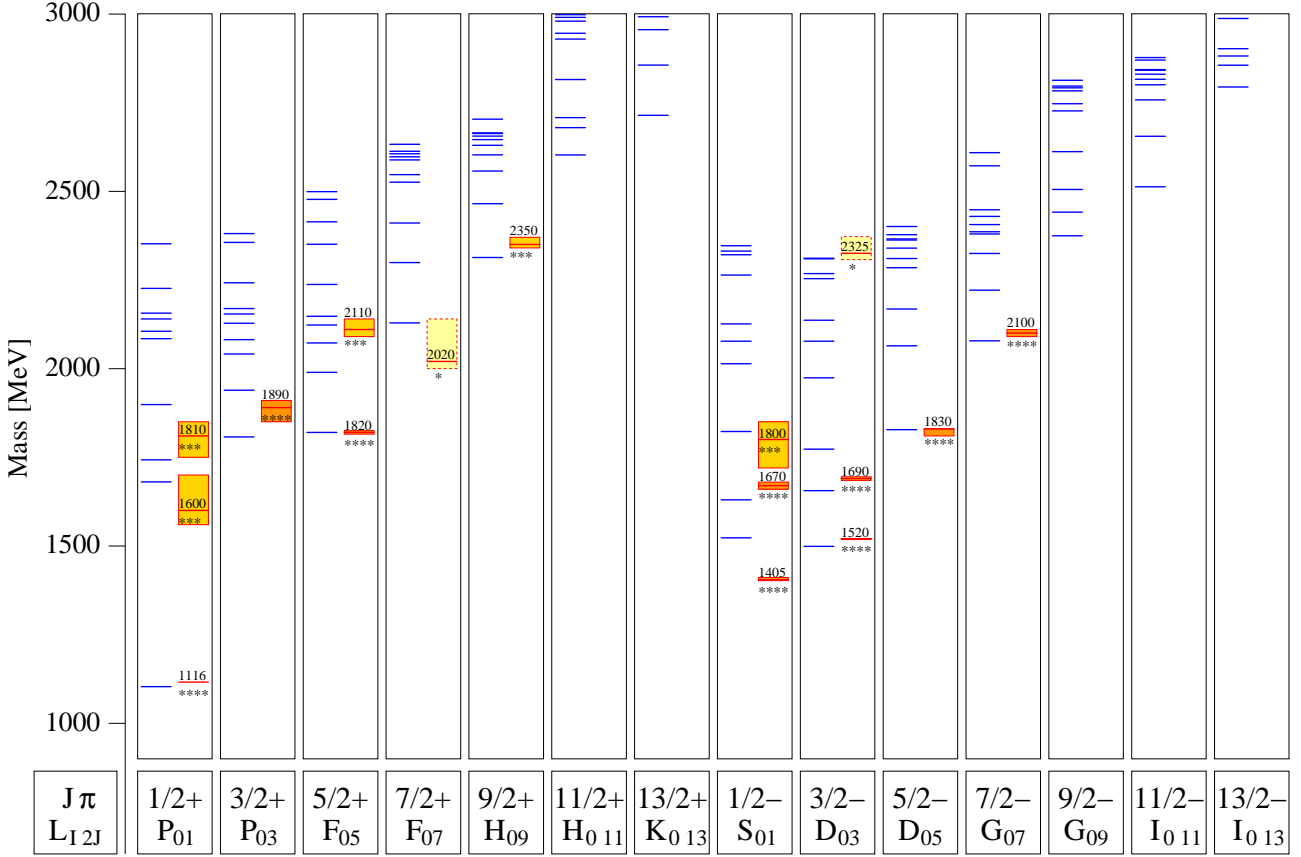}}
\caption{The left part of each column displays the calculated
    $\Lambda^*$ spectrum~\cite{loering3} for positive and negative
    parity states with spins up to $J=13/2$ and masses up to
    $3000$~MeV.  The predictions are compared with the spectrum from
    ref.~\cite{PDG2004} (right part of each column). The stars
    indicate the confidence level for the existence of each state. The
    uncertainty on a mass is indicated by the shaded area.}
\label{fig:LamM2}
\end{center}
\end{figure*}

\begin{figure}
\begin{center}
\resizebox{0.45\textwidth}{!}{\includegraphics{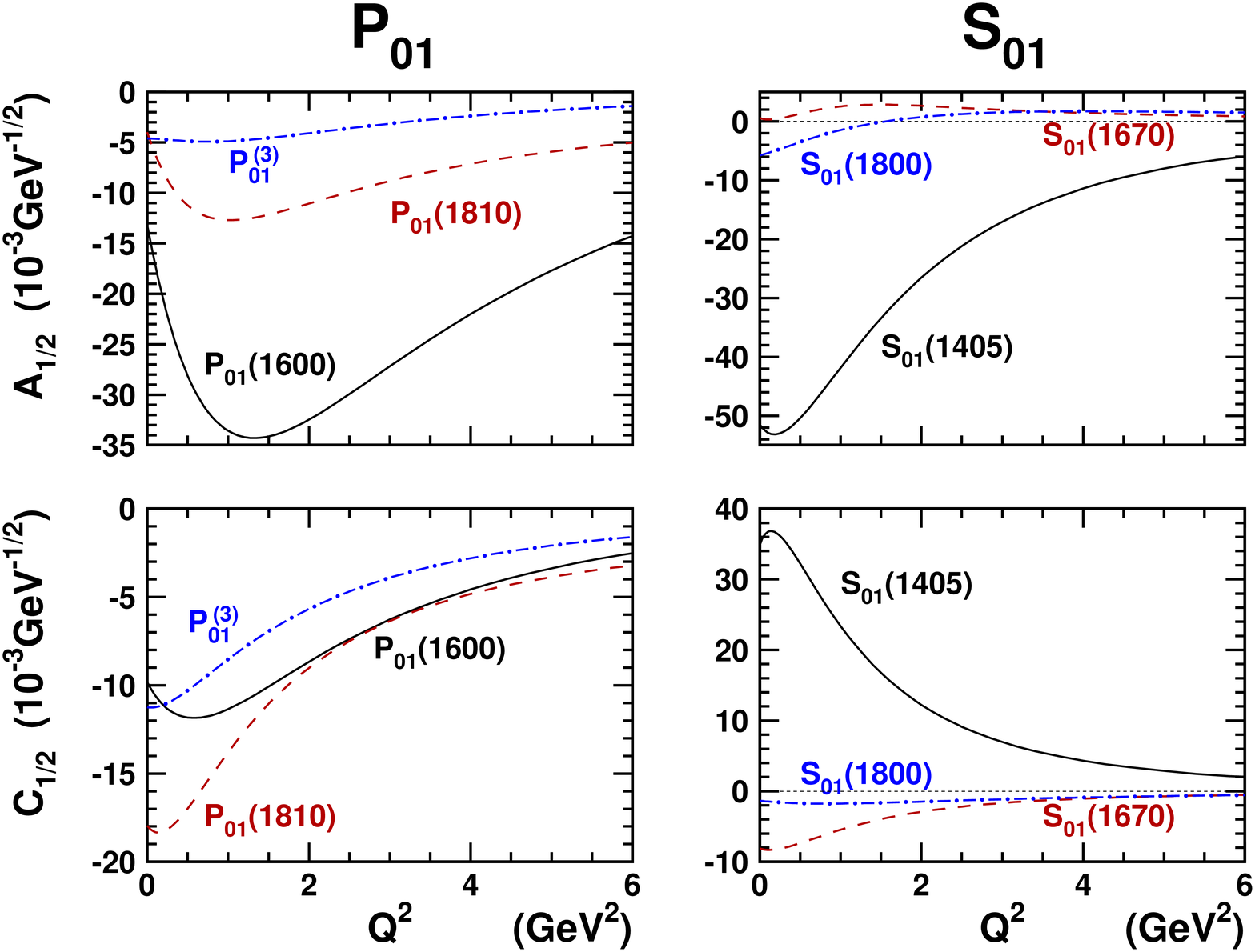}}
\caption{The $Q^2$ dependence for the $\Lambda^* + \gamma^*
    \to \Lambda$ decays for spin $J=1/2$ resonances~: left (right)
    panels show the results for the positive (negative) parity
    $\Lambda^*$ resonances.}
\label{fig:spin_1_2_iso_0}
\end{center}
\end{figure}

In this section, results are presented for the helicity amplitudes
(HA's) of $\Lambda^*$ resonances decaying to the $\Lambda$ or $\Sigma$
ground-state hyperons. The HA's are defined in
eqs.~(\ref{eq:HA_def}). We will organize our results according to the
quantum numbers of the resonances and the ground-state hyperon to
which they decay. Most of the computed low-lying states can be
identified with experimentally known resonances by comparing the
calculated with the experimental mass spectrum~\cite{loering3}. This
is illustrated in fig.~\ref{fig:LamM2} for the $\Lambda^*$
spectrum. Only for the higher-lying $F_{05}(2110)$ and $D_{03}(2325)$,
no direct correspondence with a single computed state can be made.

We will use the nomenclature adopted by the \emph{Particle Data Group}
(PDG)~\cite{PDG2004} to identify a state (\textit{e.g.} $D_{03}(1520)$
for the lowest-lying $\Lambda$ resonance with $J^\pi=3/2^-$). In those
situations where there is no clear identification possible, the
excited state with given quantum numbers will be labeled with a
number. The lowest-lying \emph{resonance} gets number '1', the second
resonance '2', \emph{etc}. Occasionally, the ground state will be
identified with a '0'. (Note that what we call a ground state, is a
member of the baryon octet.)

To illustrate the notation conventions, consider the $\Lambda^*$
spectrum in fig.~\ref{fig:LamM2}. The ground state is denoted by
$P^{(0)}_{01} \equiv P_{01}(1116)$. The first computed resonance, the
$P^{(1)}_{01}$, can be identified with the experimentally observed
$P_{01}(1600)$ resonance. For the $J^\pi = 3/2^-$ resonances, the two
lowest computed states, the $D^{(1)}_{03}$ and the $D^{(2)}_{03}$, can
be recognized as the measured $D_{03}(1520)$ and $D_{03}(1690)$
resonances respectively. The third computed resonance is as yet
unobserved experimentally, and will thus be called the
$D^{(3)}_{03}$. Note that we use the PDG conventions for denoting the
strange baryons~: $L_{I,2J}$ with the isospin $I$, spin $J$, and
$L=S,P,D$, \ldots, the orbital angular momentum of the partial wave in
which the resonance could be observed in $\Kbar \Lambda$ scattering.

\subsection{$\Lambda^* + \gamma^{(*)} \to \Lambda$ transitions}\label{sec:lam*_lam}

Our results for spin $J=1/2$, isospin $I=0$ resonances are summarized
in fig.~\ref{fig:spin_1_2_iso_0}. Already for the lowest-lying $Y^*$
resonances, one observes interesting features in the computed HA's. In
the left panel of fig.~\ref{fig:spin_1_2_iso_0}, the HA's of the three
lowest $J^\pi=\frac{1}{2}^+$ $\Lambda$ resonances are displayed. The
first excited state with the same quantum numbers as the ground-state
baryon is the analogue of the \emph{Roper resonance} in the nucleon
spectrum. In the $\Lambda$ spectrum, this state is observed
experimentally with $m \approx 1600$~MeV. For the computed state which
can be identified with the $P_{01}(1600)$ resonance, the $A_{1/2}$
amplitude reaches its maximum at $Q^2 \approx
1.5$~GeV$^2$. Accordingly, the Roper-like resonance in the $\Lambda$
spectrum may not show up in photoproduction experiments, but only in
electroproduction reactions at intermediate $Q^2$ values. Indeed, a
spacelike photon couples to the intermediate $Y^*$ resonance with a
strength proportional to its HA at that specific $Q^2$. Signals of the
$P_{01}(1600)$ resonance in electromagnetically induced kaon
production are predicted to become particularly important at $Q^2
\approx 1.5$~GeV$^2$. Another interesting feature is that the
$P_{01}(1810)$ has a relatively large $C_{1/2}$. The $C_{1/2}$
contributes to the longitudinal part of the kaon electroproduction
strength. Optimum conditions to detect signals of the $P_{01}(1810)$
are thus created when looking at the longitudinal part of the \eL
cross sections at small $Q^2$.

\begin{figure}
\begin{center}
\resizebox{0.45\textwidth}{!}{\includegraphics{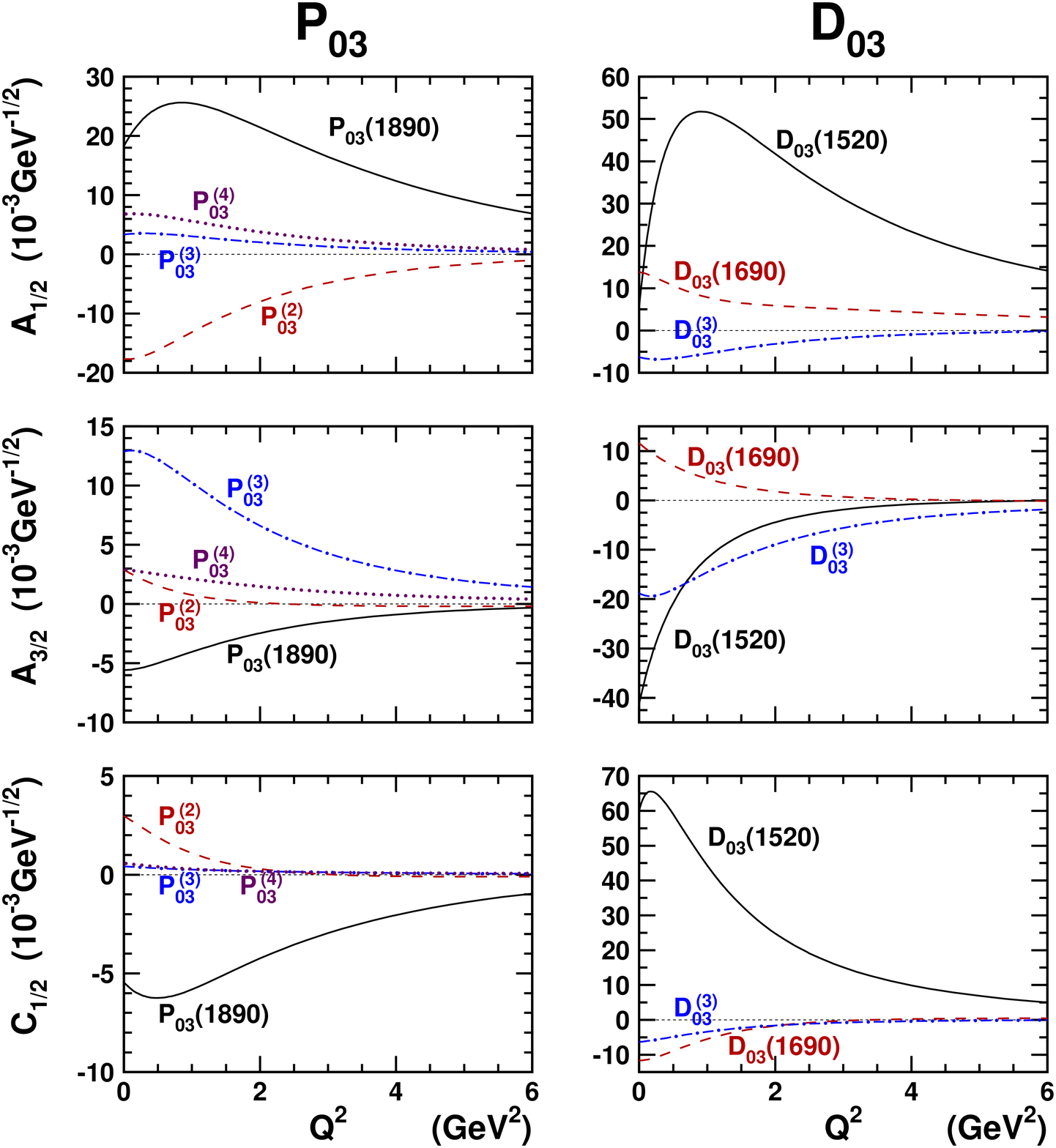}}
\caption{The $Q^2$ dependence for the $\Lambda^* + \gamma^*
    \to \Lambda$ decays for spin $J=3/2$ resonances~: left (right)
    panels show the results for the positive (negative) parity
    $\Lambda^*$ resonances.}
\label{fig:spin_3_2_iso_0}
\end{center}
\end{figure}

The most striking observation for the $S_{01}$ resonances (right
panels of fig.~\ref{fig:spin_1_2_iso_0}), is the dominance of the
lowest excitation $S_{01}(1405)$. We denote this state with the
experimental mass of the first excitation with quantum numbers
($J=1/2$, $S=-1$, $T=0$) and negative parity, but from
table~\ref{tab:PA_lam_lam} and fig.~\ref{fig:LamM2}, it is clear that
its mass is not well reproduced.  Also the calculated EM decay width
of this state is too large by a factor about $50$.  For the photon
amplitude this implies a factor of $7$, which is a huge deviation
considering the quality of our calculations for the magnetic moments
of the ground-state hyperons~\cite{tim1}. We conclude that the
$S_{01}(1405)$ is not well described in our CQ model. Possible
explanations of this discrepancy is the inadequacy of the effective
interactions used, strong rescattering effects with \textit{e.g.} the
$\Kbar N$ channel, different degrees of freedom (a three-quark
structure is possibly inadequate), a double-pole structure,
\emph{etc.}~\cite{jido_garcia-recio,garcia-recio_kolomeitsev}. For the
higher-lying $S_{01}$ resonances, our calculations predict very small
electromagnetic couplings to the $\Lambda(1116)$.

\begin{figure}
\begin{center}
\resizebox{0.45\textwidth}{!}{\includegraphics{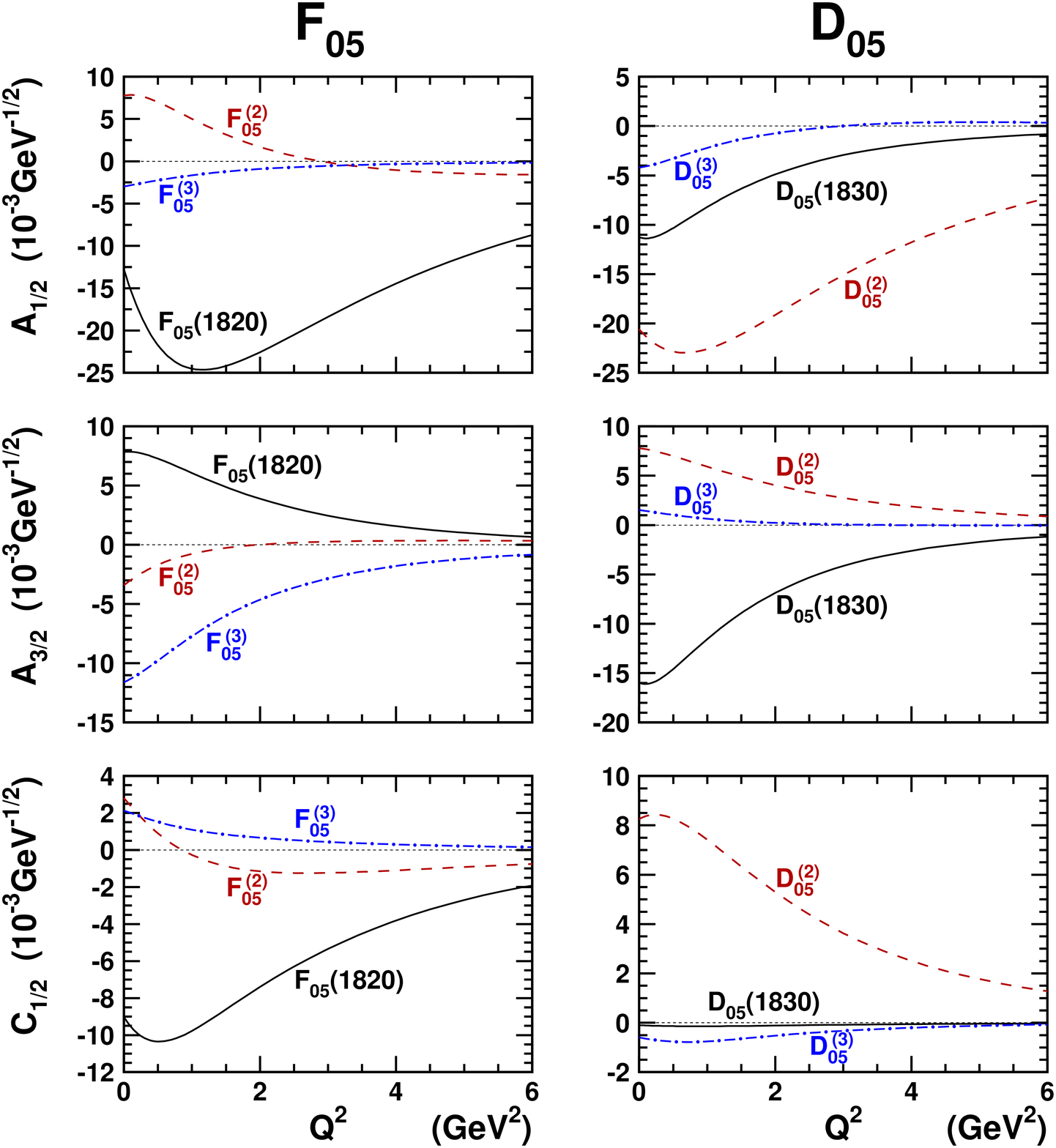}}
\caption{The $Q^2$ dependence for the $\Lambda^* + \gamma^*
    \to \Lambda$ decays for spin $J=5/2$ resonances~: left (right)
    panels show the results for the positive (negative) parity
    $\Lambda^*$ resonances.}
\label{fig:spin_5_2_iso_0}
\end{center}
\end{figure}

The HA's for the lowest-lying spin $J=3/2$ $\Lambda^*$'s are shown in
fig.~\ref{fig:spin_3_2_iso_0}. In the left panels, we consider
$P_{03}$ hyperons with positive parity. For the $A_{1/2}$, the
$P_{03}(1890)$ reaches its maximum at $Q^2 \simeq 1.0$~GeV$^2$, after
which it slowly falls to zero. The $P^{(2)}_{03}$ with a calculated
mass of $1970$~MeV (\emph{cfr.}  table~\ref{tab:PA_lam_lam}), has a
reasonably large $A_{1/2}$, but falls off rather quickly compared to
the first resonance. The other HA's, the $A_{3/2}$ and the $C_{1/2}$
are rather small for the $P_{03}$ states. Only the $P^{(3)}_{03}$ with
an expected mass of $2068$~MeV gives a modest signal in $A_{3/2}$.

The results for the $D_{03}$ helicity amplitudes are summarized in the
right panels of fig.~\ref{fig:spin_3_2_iso_0}.  Again, one notices a
peak in the $Q^2$ dependence of the first resonance at $Q^2 \simeq
0.8$~GeV$^2$ for the $A_{1/2}$, and at $Q^2 \simeq 0.2$~GeV$^2$ for
the $C_{1/2}$. Both HA's fall off slowly for large
$Q^2$-values. Systematically, we find that for almost all $I$ and $J$,
the first resonance $L^{(1)}_{I,2J}$ reaches a maximum in $A_{1/2}$
and $C_{1/2}$ at moderate values of $Q^2$. For $J \geq 3/2$, the
$A_{3/2}$'s reach their maximum at $Q^2=0$~GeV$^2$, and show a gradual
falloff with growing $Q^2$.  Furthermore, the strongest coupling is
reached at smaller values of $Q^2$ for negative-parity resonances than
for positive-parity resonances.

\begin{table}
\begin{center}
\caption{Calculated masses, photo-amplitudes and EM decay widths for
  the $\Lambda^* + \gamma \rightarrow \Lambda(1116)$
  transition. Values in between brackets denote the experimental decay
  width of the $\Lambda$ resonance as given by
  ref.~\cite{PDG2004}. Masses and decay widths are given in units of
  MeV, photo-amplitudes are given in units of $10^{-3}$~GeV$^{-1/2}$.}
\label{tab:PA_lam_lam}
\begin{tabular}{|c|c|c|c|c|}
\hline\noalign{\smallskip}
Resonance & $M_{calc}$ & $|A_{1/2}|$ & $|A_{3/2}|$ &
$\Gamma_{calc}$ \\
\noalign{\smallskip}\hline\hline\noalign{\smallskip}
$P_{01}(1600)$ & $1752$ & $13.0$ & --- & $0.104$ \\
$P_{01}(1810)$ & $1805$ & $3.97$ & --- & $0.0105$ \\
$P_{01}^{(3)}$ & $1928$ & $4.59$ & --- & $0.0174$ \\
$S_{01}(1405)$ & $1550$ & $51.5$ & --- & $0.912$ \\
 & & & & ($0.027 \pm 0.008$) \\
$S_{01}(1670)$ & $1664$ & $0.574$ & --- & $0.159 \times 10^{-3}$ \\
$S_{01}(1800)$ & $1879$ & $5.76$ & --- & $0.0252$ \\
\noalign{\smallskip}\hline\noalign{\smallskip}
$P_{03}(1890)$ & $1834$ & $18.3$ & $5.58$ & $0.129$ \\
$P_{03}^{(2)}$ & $1970$ & $17.7$ & $2.90$ & $0.142$ \\
$P_{03}^{(3)}$ & $2068$ & $3.33$ & $12.9$ & $0.0893$ \\
$P_{03}^{(4)}$ & $2116$ & $6.81$ & $2.92$ & $0.0293$ \\
$D_{03}(1520)$ & $1511$ & $5.50$ & $41.2$ & $0.258$ \\
 & & & & ($0.125^{+0.042}_{-0.038}$) \\
$D_{03}(1690)$ & $1678$ & $13.8$ & $11.6$ & $0.0815$ \\
$D_{03}^{(3)}$ & $1805$ & $6.31$ & $18.8$ & $0.130$ \\
\noalign{\smallskip}\hline\noalign{\smallskip}
$F_{05}(1820)$ & $1837$ & $12.8$ & $7.82$ & $0.0531$ \\
$F_{05}^{(2)}$ & $2012$ & $7.74$ & $3.41$ & $0.0223$ \\
$F_{05}^{(3)}$ & $2104$ & $2.98$ & $11.6$ & $0.0503$ \\
$D_{05}(1830)$ & $1843$ & $11.3$ & $16.0$ & $0.0916$ \\
$D_{05}^{(2)}$ & $2114$ & $20.6$ & $7.78$ & $0.172$ \\
$D_{05}^{(3)}$ & $2219$ & $4.22$ & $1.53$ & $0.00805$ \\
\noalign{\smallskip}\hline
\end{tabular}
\end{center}
\end{table}

The $D_{03}(1520)$ will couple quite strongly to virtual photons with
$Q^2\approx0.5$~GeV$^2$. In a partial wave analysis of \eL data, the
first $D_{03}$-resonance is likely to overwhelm the effect of the
higher-lying resonances with identical quantum numbers, which have
only moderate HA's. In table~\ref{tab:PA_lam_lam}, one notices that
for the $D_{03}(1520)$, the EM decay width is known up to a factor of
roughly two. The computed value is about $50\%$ larger than the upper
limit of the experimental width. However, the EM decay width could be
influenced by strong mixing effects with the $\Kbar N$-channel
(threshold around $1433$~MeV), which are not included in the model.

In the isobar model developed for \ggsY processes in
refs.~\cite{stijn12,stijn3,stijn4,thesisstijn}, resonances up to
$J\leq3/2$ are included. Therefore, it is instructive to see whether
there is evidence from CQ calculations to justify this
approximation. The HA's for the $J=5/2$ hyperon resonances are shown
in fig.~\ref{fig:spin_5_2_iso_0}. In the left panels, the HA's of the
three lowest-lying states with quantum numbers $J^\pi = \frac{5}{2}^+$
are displayed. Again, one observes a pronounced maximum in the
$A_{1/2}$ and $C_{1/2}$ for the $F_{05}(1820)$. If we consider the
masses in table~\ref{tab:PA_lam_lam}, it is easily seen that the
computed mass of the $F^{(2)}_{05}$ is too small for it to be
identified with the experimentally observed $F_{05}(2110)$. As a
matter of fact, from fig.~\ref{fig:LamM2}, it is clear that the third,
fourth and fifth resonance have a (computed) mass approaching the
experimentally determined value. In ref.~\cite{loering3}, it is argued
that the second resonance is actually a \emph{missing} hyperon
resonance and that the experimentally determined state around
$2110$~MeV should be associated with one of the higher-lying $F_{05}$
resonances of a CQ-model calculation. The smallness of the helicity
amplitudes displayed in fig.~\ref{fig:spin_5_2_iso_0} suggests that in
photo- and electroinduced $\Lambda$-production processes, it is
unlikely that the $F_{05}^{(2)}$ and $F_{05}^{(3)}$ $\Lambda^*$
resonances will result in strong background signals.

The right panels of fig.~\ref{fig:spin_5_2_iso_0} contain the
predictions for the three lowest-lying $D_{05}$ resonances. The first
resonance can be associated with the $D_{05}(1830)$ state from
ref.~\cite{PDG2004}. Like for the $S_{01}$ and $D_{03}$ resonances,
the $A_{1/2}$ and $C_{1/2}$ reach their maximum at low, but finite
$Q^2$ values. In contrast to the $J=1/2$ and $J=3/2$ $\Lambda^*$
resonances, the second $D_{05}$ resonance has larger HA's than the
first resonance. The PDG tables do not mention evidence for this
$D^{(2)}_{05}$ state~\cite{PDG2004}. On the basis of their computed
HA's, the $F_{05}(1820)$, the $D_{05}(1830)$ and the $D^{(2)}_{05}$
resonances can be expected to contribute sizably to the \ggsL
reaction dynamics. Therefore, prudence should be exercised when
omitting $J\geq5/2$ $\Lambda^*$ resonances in an isobar description of
the \ggsL process.

\begin{figure}
\begin{center}
\resizebox{0.45\textwidth}{!}{\includegraphics{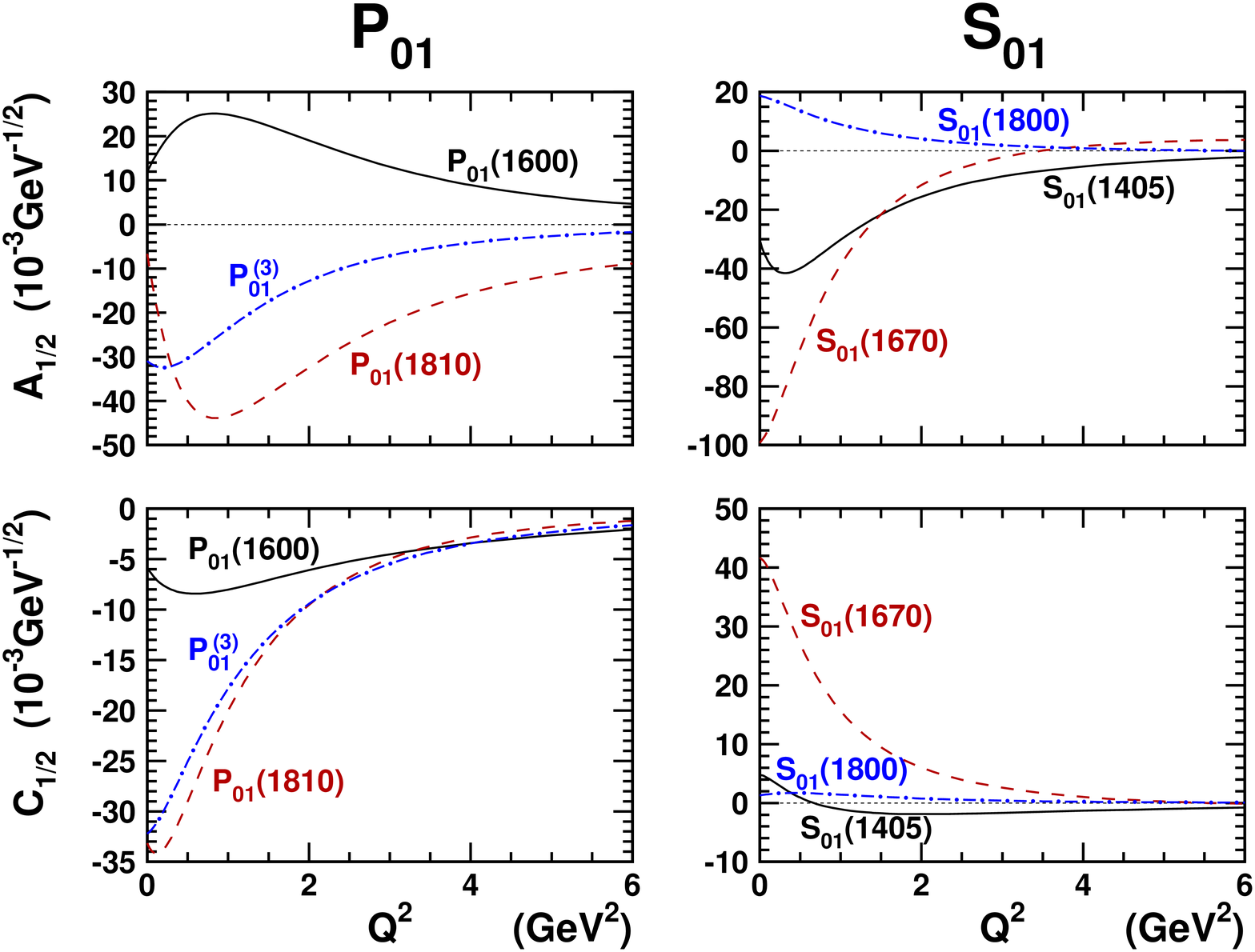}}
\caption{The $Q^2$ dependence for the $\Lambda^* + \gamma^*
    \to \Sigma$ decays for spin $J=1/2$ resonances~: left (right)
    panels show the results for the positive (negative) parity
    $\Lambda^*$ resonances.}
\label{fig:spin_1_2_iso_0_ext}
\end{center}
\end{figure}

The results for the photo-amplitudes are summarized in
table~\ref{tab:PA_lam_lam}. This table is useful for any isobar model
involving real photons which couple to a $\Lambda^*$
resonance. Experimental numbers for the EM decay of $\Lambda^*$'s are
rare. Essentially, only the decay widths for the two lowest-lying
$\Lambda$ resonances are known. Of these two, the $S_{01}(1405)$ is
often suggested to have a peculiar structure, which falls beyond the
scope of CQ-model
calculations~\cite{jido_garcia-recio,garcia-recio_kolomeitsev}.  In
view of the computed value for the EM decay width largely overshooting
the measured one, our calculations seem to confirm this
conjecture. The properties of the $D_{03}(1520)$ are, however,
reasonably well reproduced. Table~\ref{tab:PA_lam_lam} also shows that
the sole resonances for which PDG mentions an EM decay width, emerge
in our calculations indeed with the highest $\Lambda^* \to \Lambda$
widths.

More experimental information on the EM properties of the higher-lying
$\Lambda$ resonances would enable us to draw further conclusions about
the quality of our calculations. An analysis of \gsY data based on
input parameters from our CQ model would be an indirect but stringent
test of our model assumptions. At this point, we want to stress again
that we have not introduced any new parameters in the current
operator, which makes our results parameter-free predictions.

\begin{figure}
\begin{center}
\resizebox{0.45\textwidth}{!}{\includegraphics{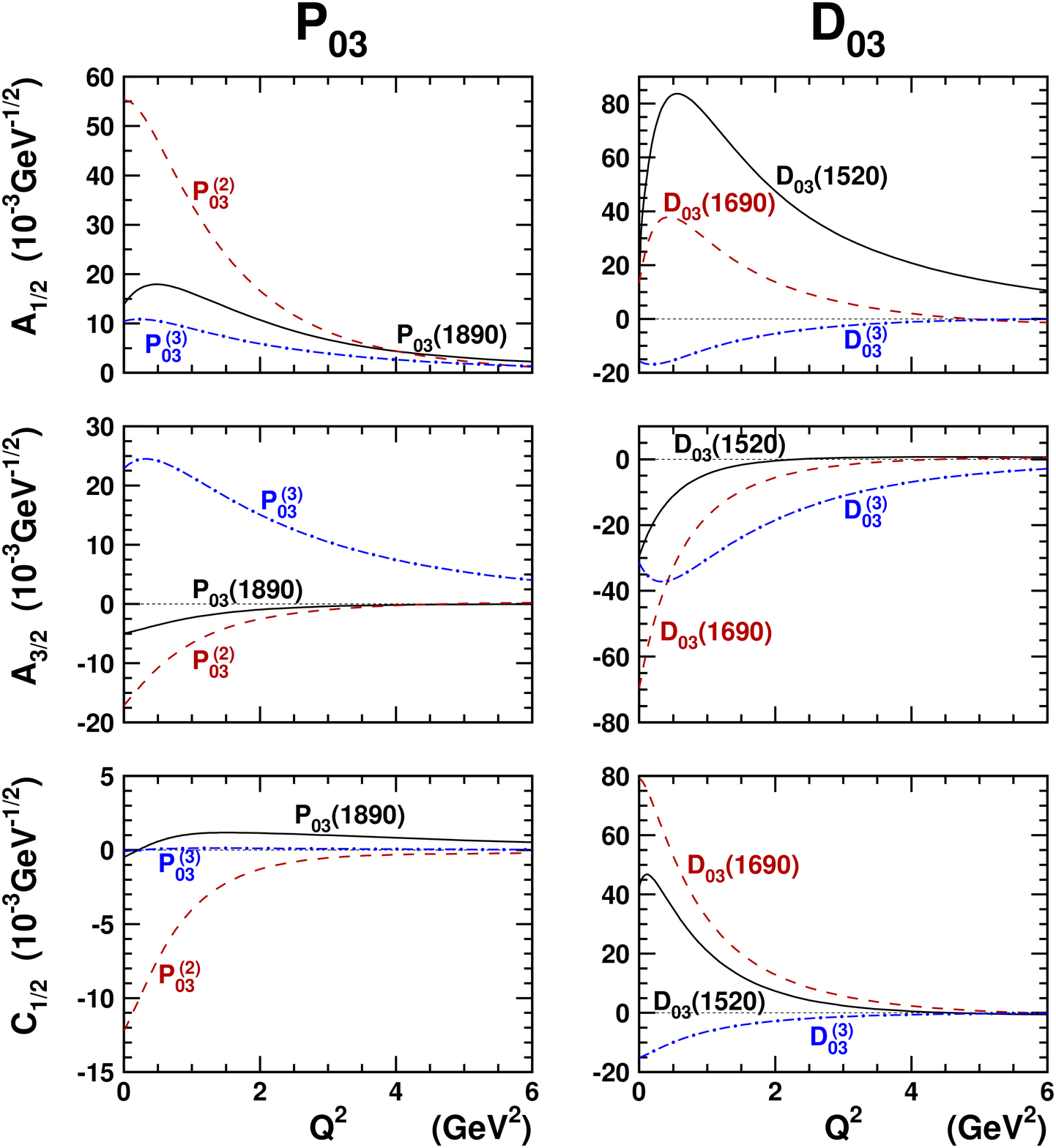}}
\caption{The $Q^2$ dependence for the $\Lambda^* + \gamma^*
    \to \Sigma$ decays for spin $J=3/2$ resonances~: left (right)
    panels show the results for the positive (negative) parity
    $\Lambda^*$ resonances.}
\label{fig:spin_3_2_iso_0_ext}
\end{center}
\end{figure}

\subsection{$\Lambda^* + \gamma^{(*)} \to \Sigma$ transitions}\label{sec:lam*_sig}

\begin{table}
\begin{center}
\caption{Calculated masses, photo-amplitudes and EM decay widths for
  the $\Lambda^* + \gamma \rightarrow \Sigma^0(1193)$ transition are
  given below. Values in between brackets denote the experimental EM
  decay width of the $\Lambda$ resonance to the $\Sigma^0(1193)$ as
  given by ref.~\cite{PDG2004}. Masses and decay widths are expressed
  in units of MeV, photo-amplitudes are given in units of
  $10^{-3}$~GeV$^{-1/2}$. Note that some masses differ from the values
  given in table~\ref{tab:PA_lam_lam}, because they were computed in a
  larger basis.}
\label{tab:PA_lam_sig}
\vspace*{10pt}
\begin{tabular}{|c|c|c|c|c|}
\hline\noalign{\smallskip}
Resonance & $M_{calc}$ & $|A_{1/2}|$ & $|A_{3/2}|$ &
$\Gamma_{calc}$ \\
\noalign{\smallskip}\hline\hline\noalign{\smallskip}
$P_{01}(1600)$ & $1713$ & $12.0$ & --- & $0.0679$ \\
$P_{01}(1810)$ & $1771$ & $6.62$ & --- & $0.0240$ \\
$P_{01}^{(3)}$ & $1928$ & $30.9$ & --- & $0.727$ \\
$S_{01}(1405)$ & $1538$ & $30.3$ & --- & $0.233$ \\
 & & & & ($0.010 \pm 0.004$/ \\
 & & & & $\ 0.023 \pm 0.007$) \\
$S_{01}(1670)$ & $1649$ & $99.2$ & --- & $3.827$ \\
$S_{01}(1800)$ & $1855$ & $18.7$ & --- & $0.231$ \\
\noalign{\smallskip}\hline\noalign{\smallskip}
$P_{03}(1890)$ & $1834$ & $13.8$ & $4.94$ & $0.068$ \\
$P_{03}^{(2)}$ & $1970$ & $55.2$ & $17.1$ & $1.367$ \\
$P_{03}^{(3)}$ & $2068$ & $10.5$ & $22.9$ & $0.303$ \\
$D_{03}(1520)$ & $1506$ & $23.3$ & $30.0$ & $0.157$ \\
 & & & & ($0.304^{+0.076}_{-0.070}$) \\
$D_{03}(1690)$ & $1668$ & $13.3$ & $70.0$ & $1.049$ \\
$D_{03}^{(3)}$ & $1790$ & $15.6$ & $31.4$ & $0.353$ \\
\noalign{\smallskip}\hline
\end{tabular}
\end{center}
\end{table}

Investigations of the $\gamma^{(*)} + p \rightarrow K^+ + \Sigma^0$
reaction in ref.~\cite{stijn3} have indicated that a proper modeling
of the background terms requires the introduction of hyperon
resonances with isospin $T=0$ \emph{and} $T=1$, \textit{i.e.} $\Lambda$
as well as $\Sigma$ resonances.

In figs.~\ref{fig:spin_1_2_iso_0_ext}
and~\ref{fig:spin_3_2_iso_0_ext}, we display the HA's for the spin
$J=1/2$ and $J=3/2$ resonances respectively. Again, one observes that
some HA's reach a maximum at moderate values for the momentum-transfer
squared ($Q^2<1.5$~GeV$^2$). This maximum is particularly pronounced
for the $A_{1/2}$ of the $P_{01}(1810)$ and $D_{03}(1520)$
resonances. For these states, the HA at its maximum is more than
double the value at $Q^2=0$.

In table~\ref{tab:PA_lam_sig}, the results for the electromagnetic
$\Lambda^*(J=1/2,3/2) \to \Sigma^0(1193)$ decays are summarized for
$Q^2=0$. The EM decay width for $S_{01}(1405) \to \Sigma^0(1193)$ is
clearly overestimated. This is similar to the $S_{01}(1405) \to
\Lambda$ result of table~\ref{tab:PA_lam_lam}, and could be attributed
to the peculiar structure of this resonance. The predicted decay width
of the $D_{03}(1520)$ is about a factor of $2$ smaller than the
experimentally determined value. This is in contrast with the
$D_{03}(1520) \to \Lambda$ transition of table~\ref{tab:PA_lam_lam},
where the width is overestimated by about a factor of $2$. The
discrepancy between computed and measured values might be attributed
to the $D_{03}(1520) \to \Kbar N \to \gamma Y$ two-step process, which
may interfere destructively with the direct $D_{03}(1520) \to \gamma
Y$ process if $Y=\Lambda$ and constructively if $Y=\Sigma^0$.

The computed EM decay width of $3.827$~MeV for the $S_{01}(1670)$ is
exceptionally large. It represents about $10\%$ of the reported total
decay width $\Gamma^{tot}_{exp}=25-50$~MeV~\cite{PDG2004}. The
\emph{Crystal Ball Collaboration} at Brookhaven has investigated
${\Kbar}^- p$ scattering up to $W \sim 1680$~MeV~\cite{prakhov01}, and
report a cross section for the ${\Kbar}^- p \to \gamma \Sigma^0$
reaction of $\sigma_{tot}=134$~${\mu}$b at a kaon lab-momentum of
$p^{lab}_K=750$~MeV ($W=1677$~MeV). This is roughly four times as
large as the cross section for the ${\Kbar}^- p \to \gamma \Lambda$
reaction ($\sigma_{tot}=31$~${\mu}$b). This observation is consistent
with the calculated EM decay width for the $S_{01}(1670) \to \Sigma^0$
transition being much larger than for the $S_{01}(1670) \to \Lambda$
process.

\begin{figure}
\begin{center}
\resizebox{0.45\textwidth}{!}{\includegraphics{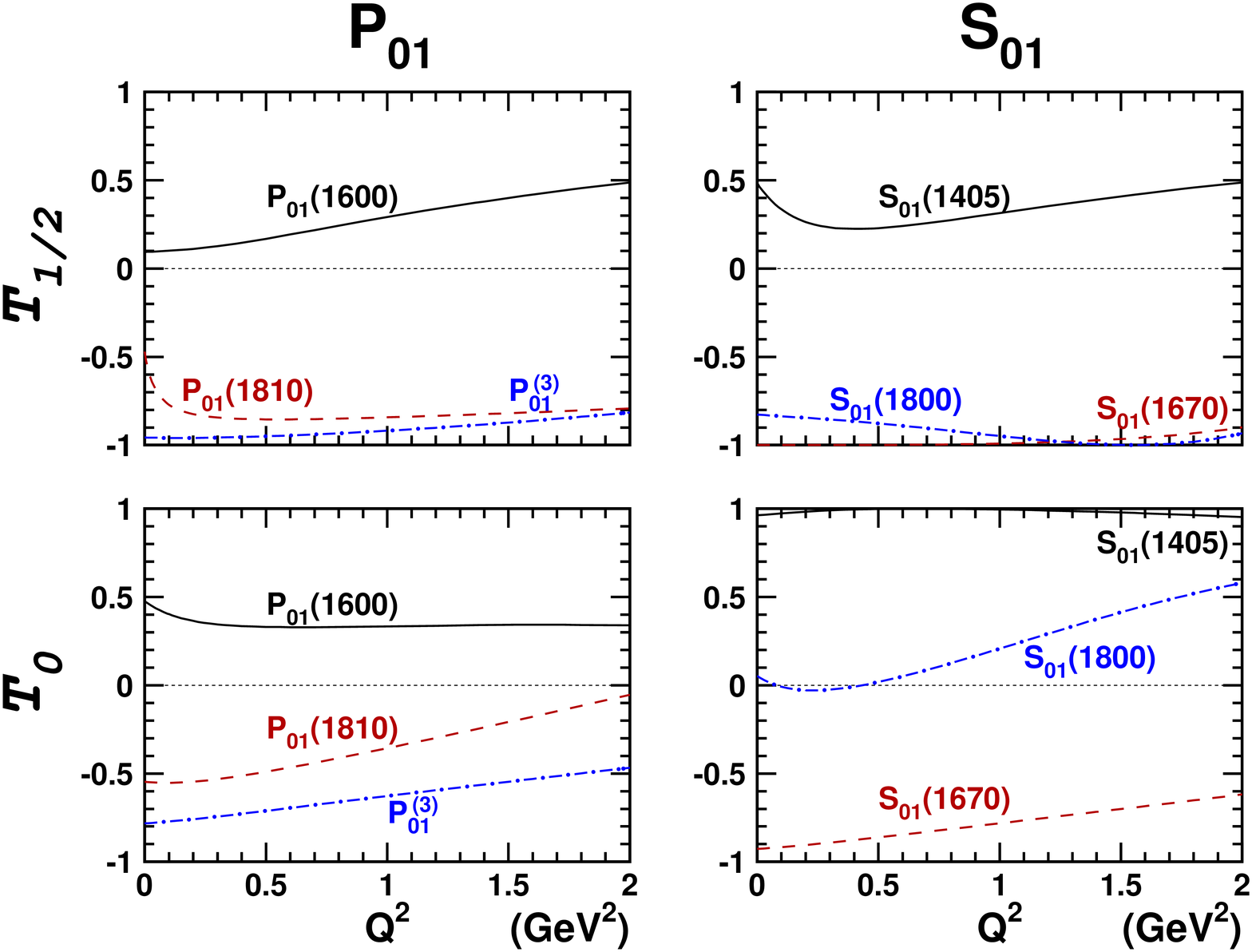}}
\caption{The asymmetries for $0 < Q^2 < 2.0$~GeV$^2$ as defined in
 eqs.~(\ref{eq:asym_sig_lam_lam*}) for the EM decays to different
 isospin channels for the three lowest-lying ($J=1/2$, $S=-1$, $T=0$)
 $\Lambda^*$ resonances with positive parity (left panels) and
 negative parity (right panels).}
\label{fig:ha_sig_lam_spin_1_2}
\end{center}
\end{figure}

The ${\Kbar}^- p \to \eta \Lambda$ cross section at an invariant mass
around $1670$~MeV was analysed in ref.~\cite{manley02}, using six
coupled channels ($\Kbar N$, $\eta \Lambda$, $\pi \Sigma$, $\pi
\Sigma^*(1385)$, $\pi \pi \Lambda$, and $\pi \pi \Sigma$). A partial
decay width of $3.6 \pm 1.4$~MeV for the $S_{01}(1670) \to \eta
\Lambda$ process was reported. This is comparable in magnitude to the
computed EM decay width in table~\ref{tab:PA_lam_sig}. Therefore,
including the $\gamma \Sigma$ channel in a coupled-channel analysis of
${\Kbar}^- p$ scattering at $p_K \approx 750$~MeV seems relevant.

\begin{figure}
\begin{center}
\resizebox{0.45\textwidth}{!}{\includegraphics{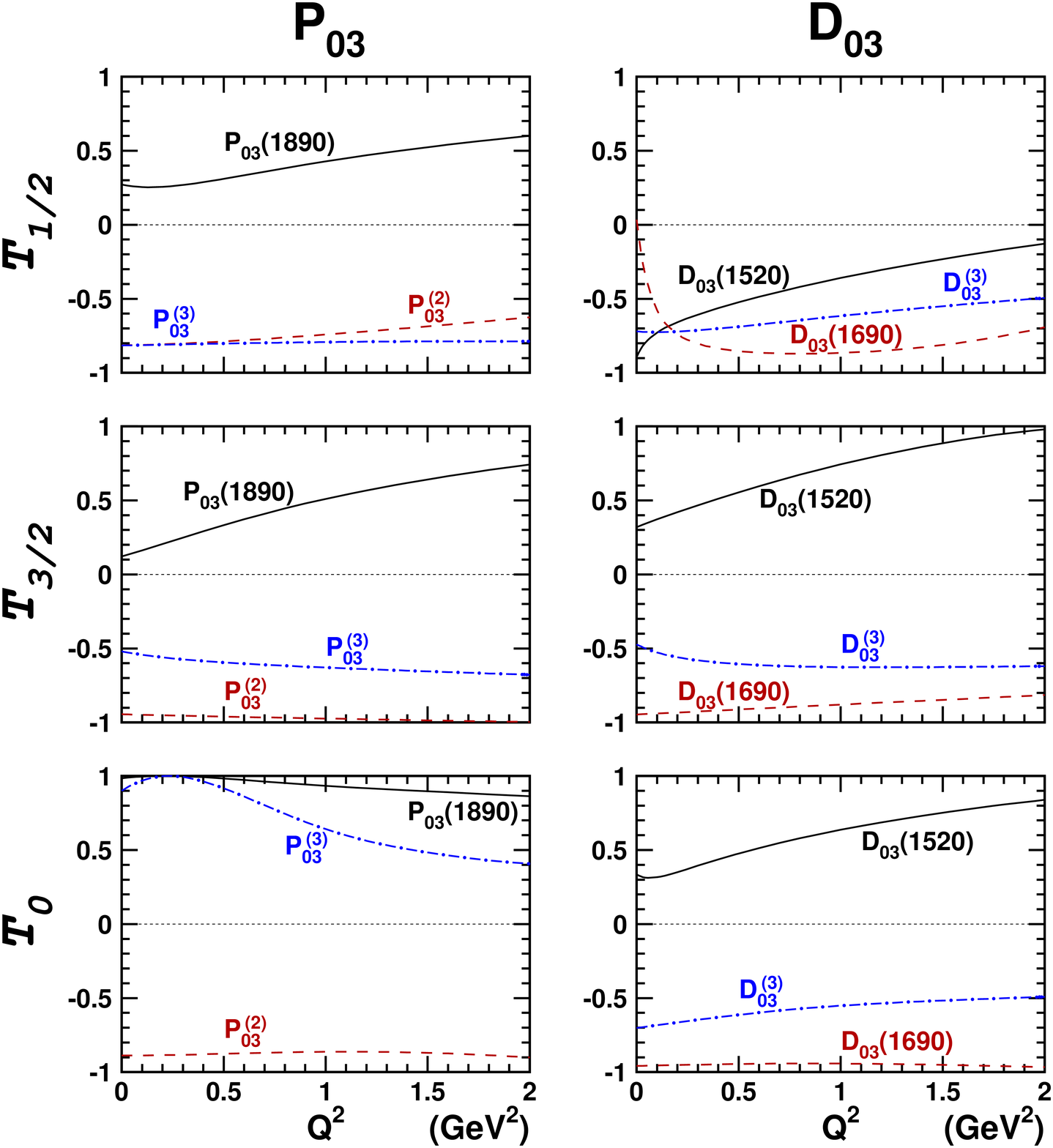}}
\caption{The asymmetries for $0 < Q^2 < 2.0$~GeV$^2$ as defined in
 eqs.~(\ref{eq:asym_sig_lam_lam*}) for the EM decays to different
 isospin channels for the three lowest-lying ($J=3/2$, $S=-1$, $T=0$)
 $\Lambda^*$ resonances with positive parity (left panels) and
 negative parity (right panels).}
\label{fig:ha_sig_lam_spin_3_2}
\end{center}
\end{figure}

For the $\Lambda^* \to \Sigma^0$ transitions, the second and the third
resonances have larger HA's than the first one. Furthermore, the
helicity amplitudes at small $Q^2$ are quite large. The difference
between $\Lambda^* \to \Lambda$ and $\Lambda^* \to \Sigma^0$ EM decays
can be made more explicit through introducing the following
\emph{isospin asymmetries}~:
\begin{subequations}
\begin{eqnarray}
\mathcal{T}_{1/2} & = & \frac {|A^{\Lambda}_{1/2}|^2 -
  |A^{\Sigma}_{1/2}|^2} {|A^{\Lambda}_{1/2}|^2 + |A^{\Sigma}_{1/2}|^2}
\; , \; \label{eq:asym_a12_sig_lam_lam*} \\
\mathcal{T}_{3/2} & = & \frac {|A^{\Lambda}_{3/2}|^2 -
  |A^{\Sigma}_{3/2}|^2} {|A^{\Lambda}_{3/2}|^2 + |A^{\Sigma}_{3/2}|^2}
\; , \; \label{eq:asym_a32_sig_lam_lam*} \\
\mathcal{T}_{0} & = & \frac {|C^{\Lambda}_{1/2}|^2 -
  |C^{\Sigma}_{1/2}|^2} {|C^{\Lambda}_{1/2}|^2 + |C^{\Sigma}_{1/2}|^2}
\; . \; \label{eq:asym_c12_sig_lam_lam*}
\end{eqnarray}
\label{eq:asym_sig_lam_lam*}
\end{subequations}
Here, the superscript $\Lambda$ ($\Sigma^0$) stands for the decay of
the resonance to the $\Lambda$ ($\Sigma^0$) ground state. It is clear
that a positive (negative) value indicates that the resonance will
preferentially decay to the $\Lambda$ ($\Sigma^0$) ground state. As
can be inferred from figs.~\ref{fig:ha_sig_lam_spin_1_2}
and~\ref{fig:ha_sig_lam_spin_3_2}, the first resonance generally has
positive isospin asymmetries, while the higher-lying resonances mostly
have negative isospin asymmetry at low momentum-transfer squared
($Q^2<2.0$~GeV$^2$).

\begin{figure}
\begin{center}
\resizebox{0.45\textwidth}{!}{\includegraphics{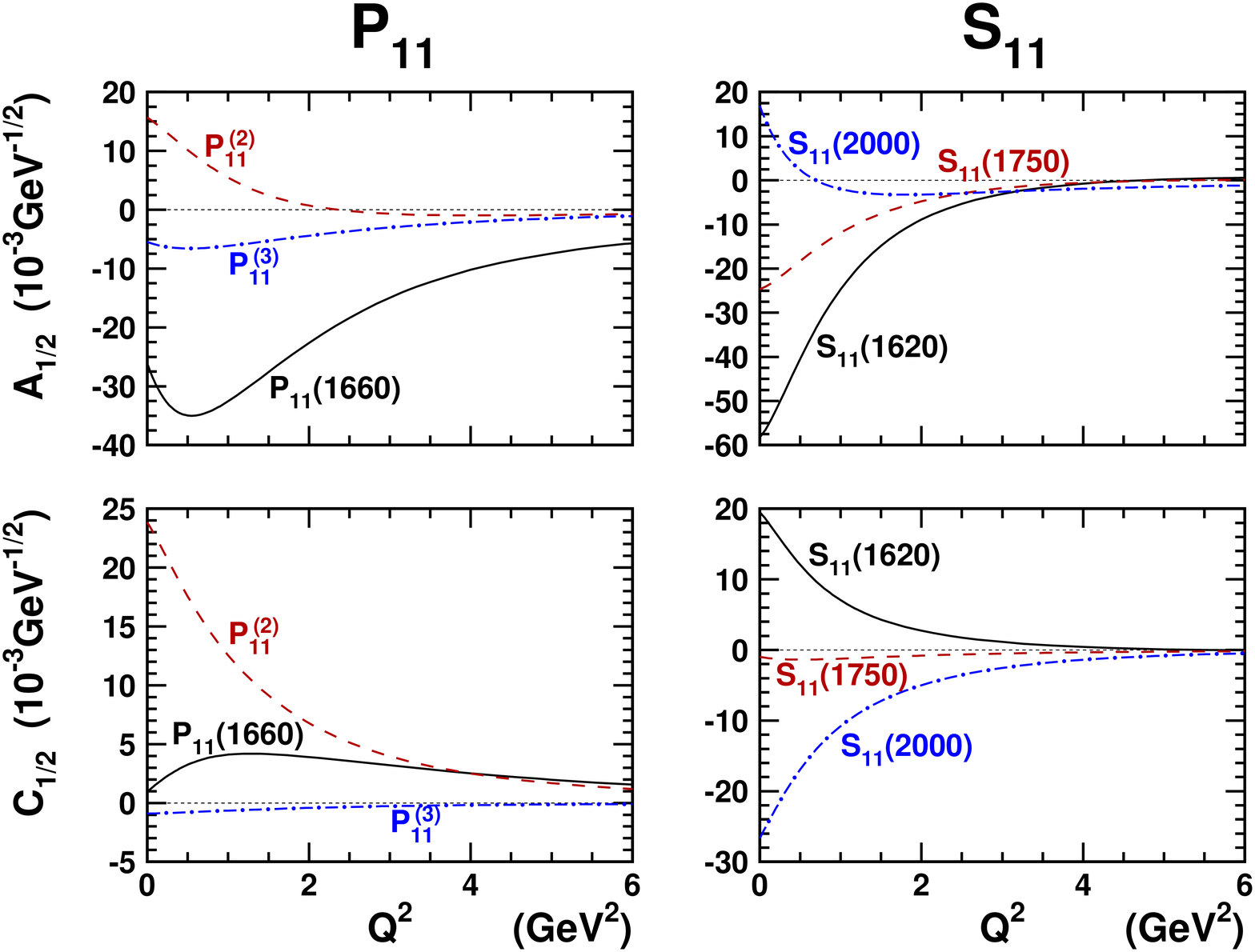}}
\caption{The $Q^2$ dependence for the ${\Sigma^*}^0 + \gamma^*
    \to \Lambda$ decays for spin $J=1/2$ resonances~: left (right)
    panels show the results for the positive (negative) parity
    $\Sigma^*$ resonances.}
\label{fig:spin_1_2_iso_1_ext}
\end{center}
\end{figure}

\section{Results for the $\Sigma$-resonances}\label{sec:sigma}

\subsection{$\Sigma^* + \gamma^{(*)} \to \Lambda$ transitions}\label{sec:sig*_lam}

In this section, we will discuss the EM helicity amplitudes of the
${\Sigma^*}^0(J=1/2,3/2) \to \Lambda$ process. The experimental
situation for the $\Sigma$ spectrum is even worse than for the
$\Lambda$. Except for the octet $\Sigma(1193)$ and the decuplet
$\Sigma^*(1385)$, only four \emph{$4$-star} and four \emph{$3$-star}
resonances are reported in ref.~\cite{PDG2004}, and of these, the spin
and parity is unknown for the $\Sigma(2250)$. Furthermore, to our
knowledge there are no data with regard to the EM properties of these
resonances.

\begin{table}
\begin{center}
\caption{Calculated masses, photo-amplitudes and EM decay widths for
  the ${\Sigma^*}^0 + \gamma \to \Lambda(1116)$ transitions for the
  lowest-lying $J=1/2$ (top rows) and $J=3/2$ (bottom rows)
  resonances. Masses and decay widths are given in units of MeV,
  photo-amplitudes are given in units of $10^{-3}$~GeV$^{-1/2}$. In
  the last column, the value in between brackets denotes the
  experimental upper limit of the EM decay width of the $\Sigma$
  resonance to the $\Lambda(1116)$ as given by ref.~\cite{PDG2004}.}
\label{tab:PA_sig_lam}
\vspace*{10pt}
\begin{tabular}{|c|c|c|c|c|}
\hline\noalign{\smallskip}
Resonance & $M_{calc}$ & $|A_{1/2}|$ & $|A_{3/2}|$ & $\Gamma_{calc}$ \\
\noalign{\smallskip}\hline\hline\noalign{\smallskip}
$P_{11}(1660)$ & $1801$ & $26.1$ & --- & $0.451$ \\
$P_{11}^{(2)}$ & $1967$ & $15.7$ & --- & $0.216$ \\
$P_{11}^{(3)}$ & $2049$ & $5.47$ & --- & $0.0294$ \\
$S_{11}(1620)$ & $1640$ & $58.2$ & --- & $1.551$ \\
$S_{11}(1750)$ & $1800$ & $24.7$ & --- & $0.403$ \\
$S_{11}(2000)$ & $1813$ & $16.9$ & --- & $0.193$ \\
\noalign{\smallskip}\hline\noalign{\smallskip}
$P_{13}(1385)$ & $1409$ & $63.7$ & $109.8$ & $1.527$ \\
 & & & & ($< 13.94$) \\
$P_{13}(1840)$ & $1902$ & $-29.4$ & $9.56$ & $0.378$ \\
$P_{13}(2080)$ & $1950$ & $26.6$ & $44.7$ & $1.155$ \\
$D_{13}(1580)$ & $1675$ & $14.2$ & $-36.8$ & $0.390$ \\
$D_{13}(1670)$ & $1727$ & $36.1$ & $61.9$ & $1.457$ \\
$D_{13}^{(3)}$ & $1780$ & $-38.1$ & $-27.8$ & $0.706$ \\
\noalign{\smallskip}\hline
\end{tabular}
\end{center}
\end{table}

The predictions from the Bethe-Salpeter model for the photo-amplitudes
and EM decay widths are presented in table~\ref{tab:PA_sig_lam} for
the $J=1/2$ (top rows) and $J=3/2$ (bottom rows) $\Sigma^*$
resonances. The three lowest $\Sigma^*$'s with $J=1/2$ from our
calculations are referred to as $P_{11}(1660)$, $P^{(2)}_{11}$ and
$P^{(3)}_{11}$. The existence of the $P_{11}(1770)$ is based on one
analysis, and is questionable~\cite{PDG2004}. Therefore, it is argued
in ref.~\cite{loering3}, that the $P_{11}(1770)$ should be
disregarded, and that the $P_{11}(1880)$ is actually the second-lowest
resonance $P^{(2)}_{11}$. Even then, the predicted masses are about
$100$~MeV too high. For the negative-parity $\Sigma^*$ resonances, the
situation for the $J=1/2$ resonances is more clear. The identification
of the two lowest-lying computed states with the experimentally
observed ones is straightforward by comparing the measured and the
predicted masses. The $S_{11}(2000)$ can be identified with the third
computed state, since the value of $2000$~MeV for its experimental
mass is a very rough estimate~\cite{PDG2004}. The computed EM decay
widths in table~\ref{tab:PA_sig_lam} decrease with increasing mass for
the $P_{11}$ as well as for the $S_{11}$ resonances.

Table~\ref{tab:PA_sig_lam} also shows the EM decay widths of the
lowest lying $J=3/2$ $\Sigma$ resonances. One clearly observes rather
large values for the $P_{13}(1385)$, the $P_{13}(2080)$ and the
$D_{13}(1670)$ resonances. The first is a member of the baryon
decuplet. The PDG provides only a rough upper limit around $13.94$~MeV
for the EM decay width of the $\Sigma^*(1385)$ to the
$\Lambda(1116)$~\cite{PDG2004}. Our computed value is well below that
limit. The $D_{13}(1670)$ resonance could magnify the effect of the
$S_{01}(1670)$ in the ${\Kbar}^- p \to \gamma \Lambda$ process,
increasing the total cross section of the latter reaction even more at
$W \approx 1670$~MeV.

For the $P_{11}$ resonances, the results for the HA's are displayed in
the left panels of fig.~\ref{fig:spin_1_2_iso_1_ext}. The
$P_{11}(1660)$, which is the analogue of the Roper resonance in the
$\Sigma$ spectrum, has the largest $A_{1/2}$, reaching a maximum at
$Q^2 \approx 0.5$~GeV$^2$. The second resonance has the largest
$C_{1/2}$ for small to moderate $Q^2$-values. The HA's for the
$P_{11}$ $\Sigma^* \to \Lambda$ decays are comparable to those for the
$P_{01}$ $\Lambda^* \to \Lambda$ decays. Therefore, the $\Sigma^*$'s
can be expected to contribute significantly to the background of the
\ggsL and the ${\Kbar}^- p \to \gamma \Lambda$ cross sections. This
observation is even more relevant to the $J^\pi=1/2^-$ resonances, for
which the HA's are depicted in the right panels of
fig.~\ref{fig:spin_1_2_iso_1_ext}. One observes a large $A_{1/2}$ for
the $S_{11}(1620)$, which is a \emph{$2$-star} resonance in
ref.~\cite{PDG2004}. This is also clear from the large EM decay width
of this resonance, reported in table~\ref{tab:PA_sig_lam}. However,
the data for the $\Kbar^- p \to \gamma \Lambda$
process~\cite{phaisangittisakul01}, do not show a significant
enhancement at $W \approx 1620$~MeV ($p_K \approx 629$~MeV). This
could be explained by a small coupling of the $S_{11}(1620)$ resonance
to the $\Kbar N$ channel.

\begin{figure}
\begin{center}
\resizebox{0.45\textwidth}{!}{\includegraphics{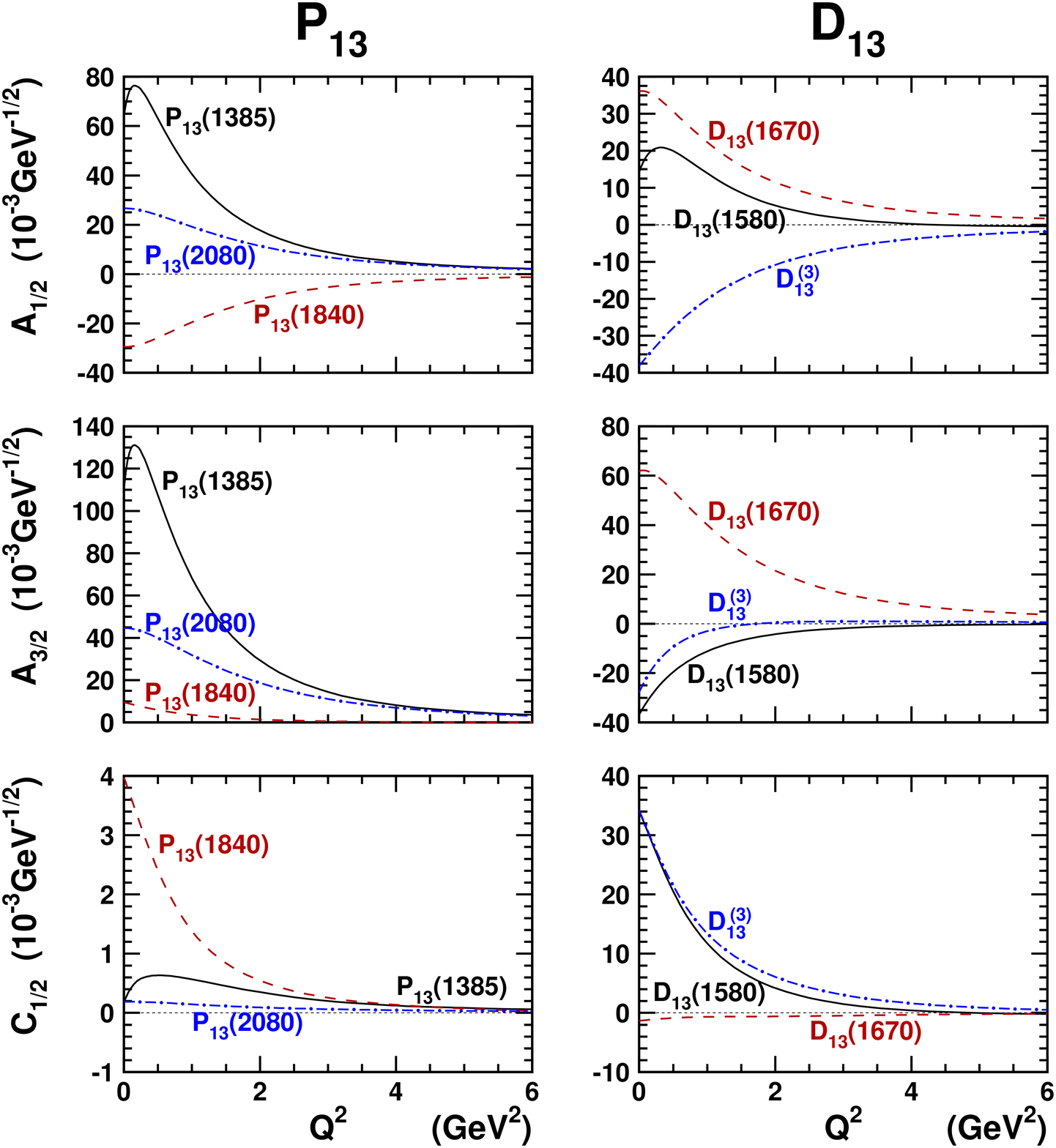}}
\caption{The $Q^2$ dependence for the ${\Sigma^*}^0 + \gamma^*
    \to \Lambda$ decays for spin $J=3/2$ resonances~: left (right)
    panels show the results for the positive (negative) parity
    $\Sigma^*$ resonances.}
\label{fig:spin_3_2_iso_1_ext}
\end{center}
\end{figure}

The computed helicity amplitudes for the $J=3/2$ $\Sigma^*$'s are
displayed in fig.~\ref{fig:spin_3_2_iso_1_ext}. The decuplet member
$P_{13}(1385)$ has the largest HA's of the positive-parity
resonances. The $C_{1/2}$'s of the three lowest $P_{13}$ states are
all rather small. The helicity amplitudes of the three lowest-lying
negative-parity $\Sigma$ resonances are moderate, except for the the
$C_{1/2}$ of the $D_{13}(1670)$, which practically vanishes.

\subsection{$\Sigma^* + \gamma^{(*)} \to \Sigma$ transitions}\label{sec:sig*_sig}

\begin{figure}
\begin{center}
\resizebox{0.45\textwidth}{!}{\includegraphics{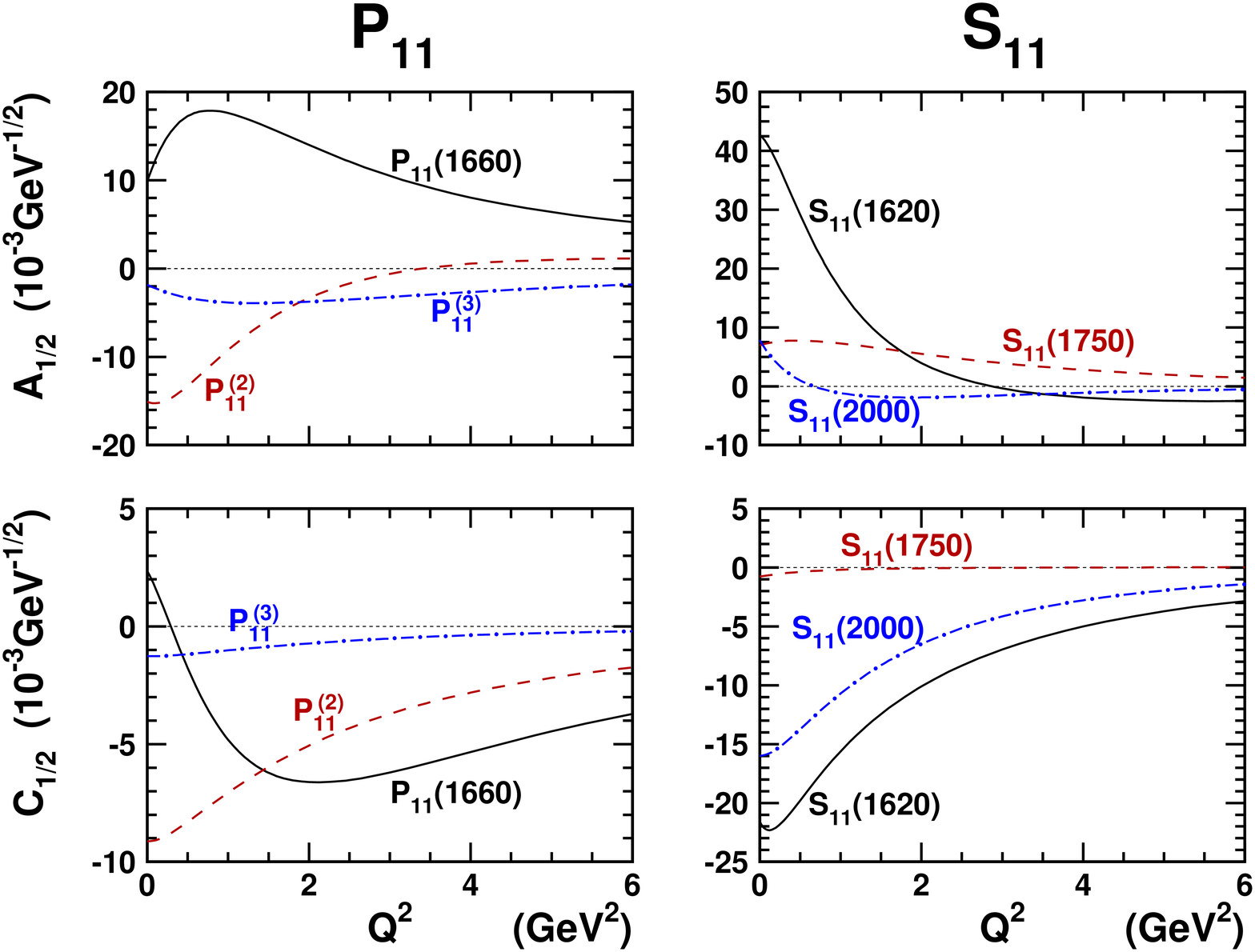}}
\caption{The $Q^2$ dependence for the ${\Sigma^*}^0 + \gamma^*
    \to \Sigma^0$ decays for spin $J=1/2$ resonances~: left (right)
    panels show the results for the positive (negative) parity
    $\Sigma^*$ resonances.}
\label{fig:spin_1_2_iso_1_n}
\end{center}
\end{figure}

\begin{table}
\begin{center}
\caption{Calculated masses, photo-amplitudes and EM decay widths for
  the $\Sigma^* + \gamma \rightarrow \Sigma(1193)$ transitions for the
  $J=1/2$ $\Sigma^*$ resonances. The charge of the $\Sigma^*$
  isospin-triplet member is indicated by the superscript
  $^{(0,+,-)}$. Masses and decay widths are given in units of MeV,
  photo-amplitudes are given in units of $10^{-3}$~GeV$^{-1/2}$.}
\label{tab:PA_sig_sig}
\vspace*{10pt}
\begin{tabular}{|c|c|c|c|c|}
\hline\noalign{\smallskip}
Resonance & $M_{calc}$ & $|A_{1/2}|$ & $\Gamma_{calc}$ \\
\noalign{\smallskip}\hline\hline\noalign{\smallskip}
$P^0_{11}(1660)$ & $1801$ & $9.91$ & $0.0578$ \\
$P^{(2) 0}_{11}$ & $1967$ & $15.1$ & $0.186$ \\
$P^{(3) 0}_{11}$ & $2049$ & $1.83$ & $0.00311$ \\
$S^0_{11}(1620)$ & $1640$ & $42.7$ & $0.688$ \\
$S^0_{11}(1750)$ & $1800$ & $6.96$ & $0.0284$ \\
$S^0_{11}(2000)$ & $1813$ & $7.86$ & $0.0373$ \\
\noalign{\smallskip}\hline\noalign{\smallskip}
$P^+_{11}(1660)$ & $1801$ & $35.3$ & $0.733$ \\
$P^{(2) +}_{11}$ & $1967$ & $54.8$ & $2.446$ \\
$P^{(3) +}_{11}$ & $2049$ & $4.86$ & $0.0219$ \\
$S^+_{11}(1620)$ & $1640$ & $125.6$ & $5.955$ \\
$S^+_{11}(1750)$ & $1800$ & $4.80$ & $0.0135$ \\
$S^+_{11}(2000)$ & $1813$ & $10.3$ & $0.0641$ \\
\noalign{\smallskip}\hline\noalign{\smallskip}
$P^-_{11}(1660)$ & $1801$ & $15.5$ & $0.141$ \\
$P^{(2) -}_{11}$ & $1967$ & $24.6$ & $0.493$ \\
$P^{(3) -}_{11}$ & $2049$ & $1.20$ & $0.00136$ \\
$S^-_{11}(1620)$ & $1640$ & $40.3$ & $0.613$ \\
$S^-_{11}(1750)$ & $1800$ & $9.12$ & $0.0488$ \\
$S^-_{11}(2000)$ & $1813$ & $5.41$ & $0.0177$ \\
\noalign{\smallskip}\hline
\end{tabular}
\end{center}
\end{table}

The $\Sigma^* + \gamma^{(*)} \to \Sigma(1193)$ process differs from
the ones of previous sections in that it comes in three versions, one
for each member of the $\Sigma^*$ isospin triplet. Their EM properties
are not independent, however, because of the presumed isospin symmetry
of the interactions in the Bonn model ($u$- and $d$-quark have the
same mass and the effective interactions do not depend on the third
component of the isospin quantum number $T_z$ of the quark). Knowledge
of the helicity amplitudes for the ${\Sigma^*}^+$ and the
${\Sigma^*}^-$ allows one to obtain those for the ${\Sigma^*}^0$,
simply by taking the average. In the following, results for all three
isospin-triplet members will be presented. The charge of the particle
will be denoted in the superscript.

In fig.~\ref{fig:spin_1_2_iso_1_n}, the HA's for the ${\Sigma^*}^0 \to
\Sigma^0$ decays are displayed for the lowest-lying spin $J=1/2$
resonances. Obviously, the HA's for the $P^0_{11}$ resonances are
relatively small. This is reflected in the rather small values for the
computed EM decay widths of the $P^0_{11}$ resonances given in
table~\ref{tab:PA_sig_sig}. A larger EM response is seen for the
negative-parity states, where the $S^0_{11}(1620)$ has HA's of similar
magnitude as the ones for the decay to the $\Lambda(1116)$
(fig.~\ref{fig:spin_1_2_iso_1_ext}). The other $S^0_{11}$ resonances
have rather small HA's.

\begin{figure}
\begin{center}
\resizebox{0.45\textwidth}{!}{\includegraphics{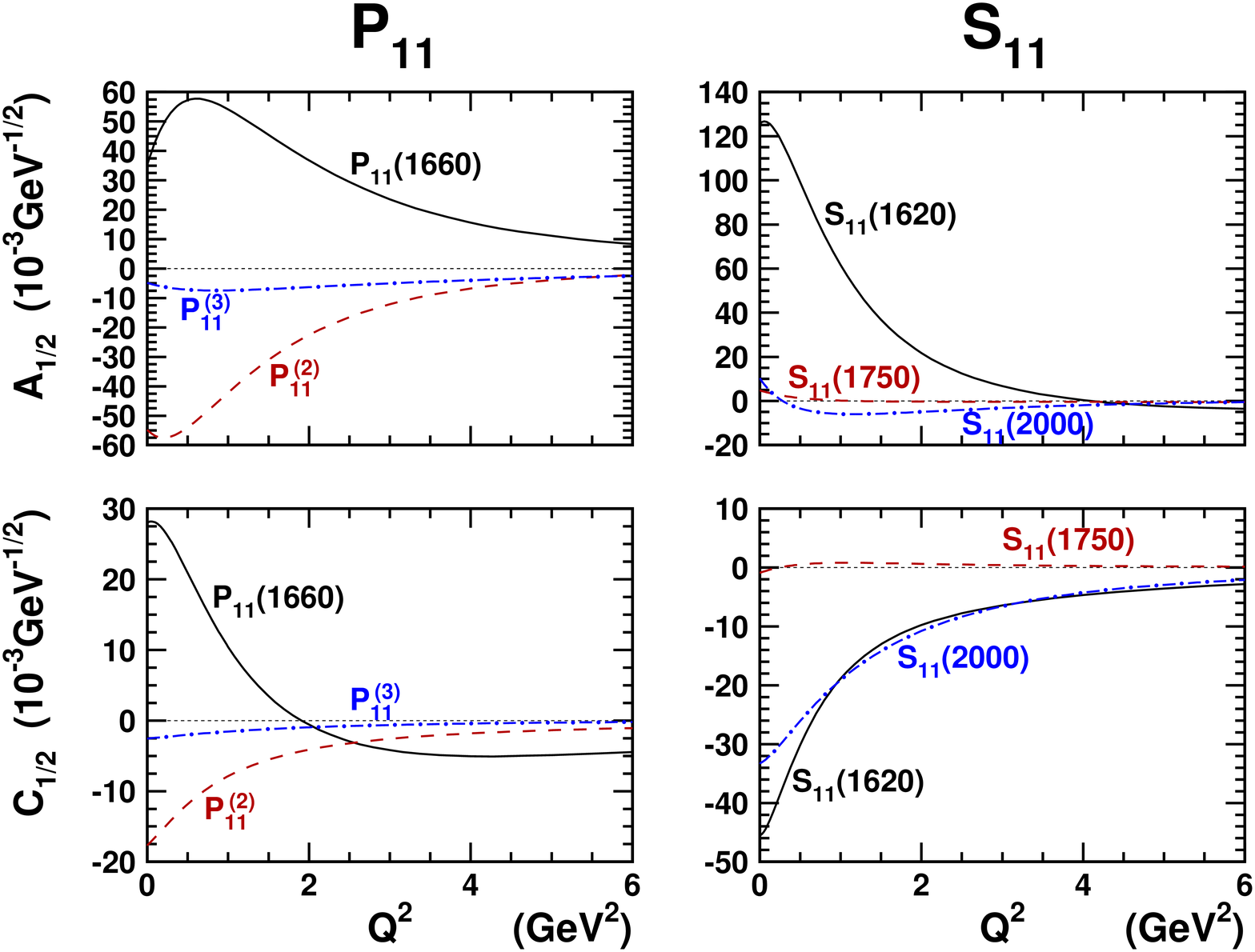}}
\caption{The $Q^2$ dependence for the ${\Sigma^*}^+ + \gamma^*
    \to \Sigma^+$ decays for spin $J=1/2$ resonances~: left (right)
    panels show the results for the positive (negative) parity
    $\Sigma^*$ resonances.}
\label{fig:spin_1_2_iso_1_p}
\end{center}
\end{figure}

The results for the positively charged members of the
$\Sigma^*$-triplets, which are presented in
fig.~\ref{fig:spin_1_2_iso_1_p}, are quite surprising. In contrast
with their neutral counterparts, the first and second $P^+_{11}$
resonances have large helicity amplitudes. This can also be deduced
from the predictions for the EM decay widths in
table~\ref{tab:PA_sig_sig}. These findings have serious implications
when modeling the background contributions in \ggsY processes. When
$Y$ is a neutral hyperon ($\Lambda$ or $\Sigma^0$), the exchanged
particle in the $u$-channel (fig.~\ref{diag:u_chan_tree}) would
necessarily be neutral. The $P^0_{11}$ resonances are likely to have a
negligible effect because of their small EM couplings. When
$Y=\Sigma^+$, the intermediate hyperon would be positively charged,
and the $P^+_{11}$ resonances could contribute sizably to the
background.

For the $S^+_{11}$ resonances, a striking feature is the large EM
decay width of the first resonance in
table~\ref{tab:PA_sig_sig}. Again, this indicates the large coupling
of the $S^+_{11}(1620)$ to the $\gamma Y$ decay channels. Furthermore,
the EM decay width $\Gamma^{EM}_{calc} \simeq 6$~MeV seems to be a
significant fraction of the poorly known total decay width
$\Gamma^{tot}_{exp}=10-106$~MeV~\cite{PDG2004}. Since the latter was
extracted from meson-baryon scattering experiments, it is possible
that the experimental status of this resonance can be improved
considerably by investigating radiative processes. The computed HA's
of the other $S^+_{11}$ resonances again turn out to be relatively
small.

\begin{figure}
\begin{center}
\resizebox{0.45\textwidth}{!}{\includegraphics{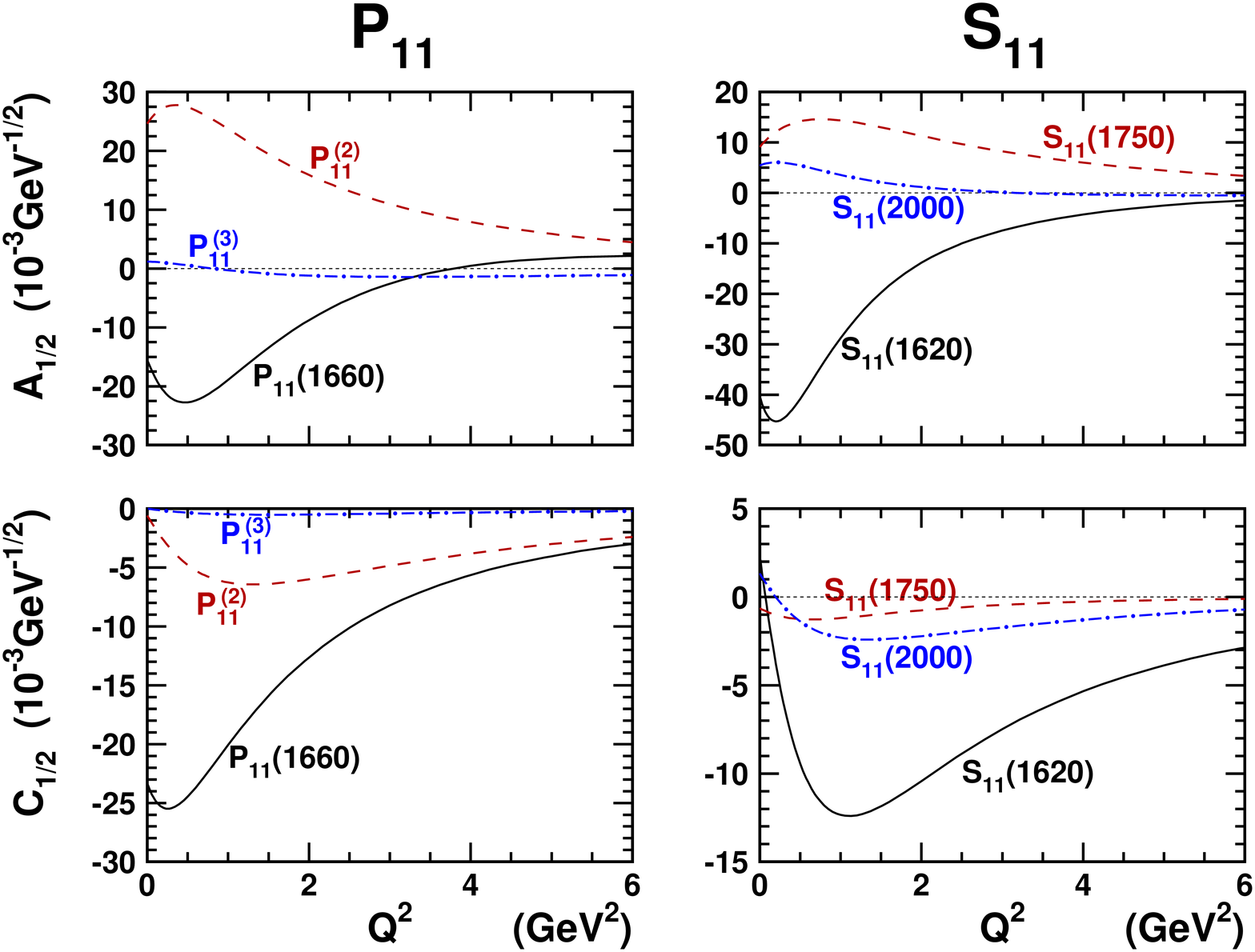}}
\caption{The $Q^2$ dependence for the ${\Sigma^*}^- + \gamma^*
    \to \Sigma^-$ decays for spin $J=1/2$ resonances~: left (right)
    panels show the results for the positive (negative) parity
    $\Lambda^*$ resonances.}
\label{fig:spin_1_2_iso_1_m}
\end{center}
\end{figure}

The calculated HA's for the $P^-_{11}$ and $S^-_{11}$ $\Sigma^*$
resonances are displayed in fig.~\ref{fig:spin_1_2_iso_1_m}. Moderate
HA's and EM decay widths (table~\ref{tab:PA_sig_sig}) are observed for
the positive-parity resonances. For the negative-parity resonances,
one notices the large $A_{1/2}$ for the $S^-_{11}(1620)$ resonance.

\begin{figure}
\begin{center}
\resizebox{0.45\textwidth}{!}{\includegraphics{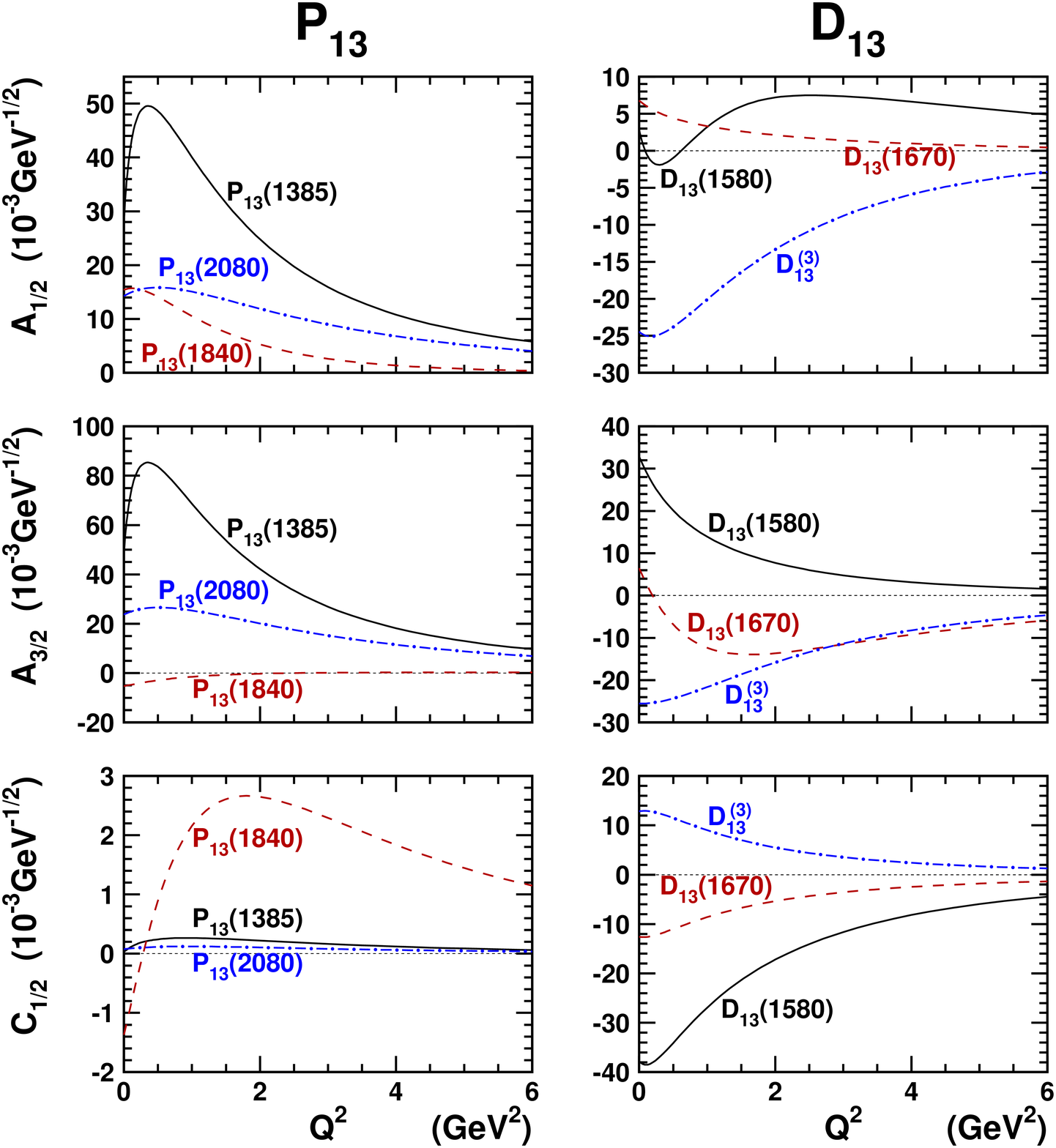}}
\caption{The $Q^2$ dependence for the ${\Sigma^*}^0 + \gamma^*
    \to \Sigma^0$ decays for spin $J=3/2$ resonances~: left (right)
    panels show the results for the positive (negative) parity
    $\Sigma^*$ resonances.}
\label{fig:spin_3_2_iso_1_n}
\end{center}
\end{figure}

The HA's for the neutral process ${\Sigma^*}^0 + \gamma^* \to
\Sigma^0$ are shown in fig.~\ref{fig:spin_3_2_iso_1_n} for $J=3/2$
resonances. For the $P_{13}$ resonances, one can point to the
relatively large $A_{1/2}$ and $A_{3/2}$ amplitudes for the decuplet
member $P_{13}(1385)$. Yet, due to the small phase space, this does
not result in a large EM decay width, as presented in
table~\ref{tab:PA_sig_sig_3_2}. The $P_{13}(2080)$ resonance, on the
other hand, has only moderately large helicity amplitudes, yet has a
larger EM decay width than the $P_{13}(1385)$ due to its larger
mass. The results for the negative-parity resonances are displayed in
the right panels of fig.~\ref{fig:spin_3_2_iso_1_n}. There, one
notices the small HA's of the $D_{13}(1670)$ resonance. Consequently,
this resonance has a small EM decay width (\emph{cfr.}
table~\ref{tab:PA_sig_sig_3_2}). The computed HA's of the other two
$D_{13}$ resonances are of intermediate magnitude.

\begin{figure}
\begin{center}
\resizebox{0.45\textwidth}{!}{\includegraphics{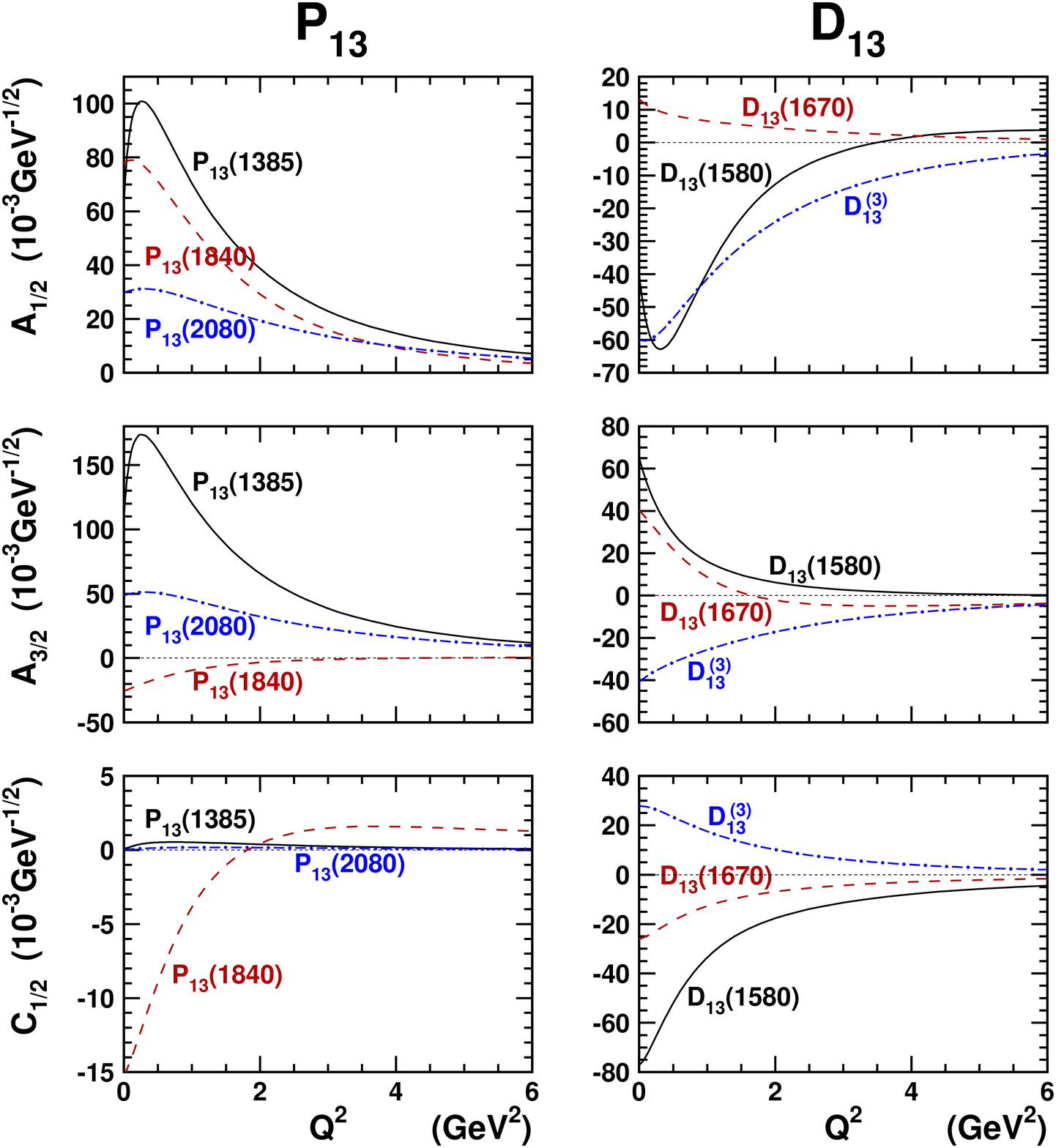}}
\caption{The $Q^2$ dependence for the ${\Sigma^*}^+ + \gamma^*
    \to \Sigma^+$ decays for spin $J=3/2$ resonances~: left (right)
    panels show the results for the positive (negative) parity
    $\Sigma^*$ resonances.}
\label{fig:spin_3_2_iso_1_p}
\end{center}
\end{figure}

\begin{figure}
\begin{center}
\resizebox{0.45\textwidth}{!}{\includegraphics{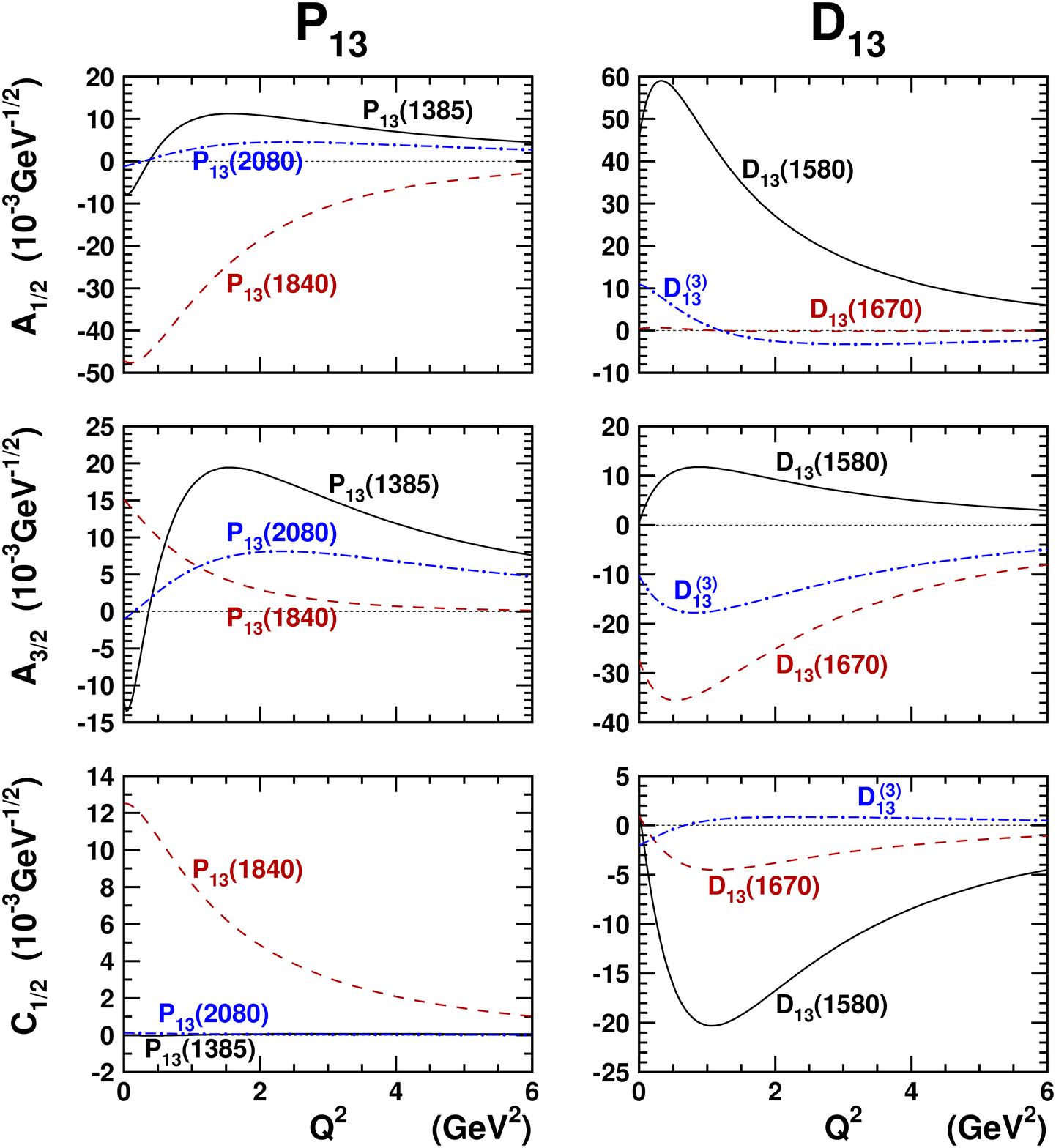}}
\caption{The $Q^2$ dependence for the ${\Sigma^*}^- + \gamma^*
    \to \Sigma^-$ decays for spin $J=3/2$ resonances~: left (right)
    panels show the results for the positive (negative) parity
    $\Lambda^*$ resonances.}
\label{fig:spin_3_2_iso_1_m}
\end{center}
\end{figure}

We also present the results for the HA's of the charged ${\Sigma^*}^+
+ \gamma^* \to \Sigma^+$ process in
fig.~\ref{fig:spin_3_2_iso_1_p}. This figure shows that resonances for
which the HA's of the neutral process were small or moderate, can
still have large HA's for the (positively-)charged process, as was the
case with the $J=1/2$ $\Sigma$ resonances. This is made even more
clear in table~\ref{tab:PA_sig_sig_3_2}, where it is seen that the EM
decay widths of the positively-charged $\Sigma^*$'s are a factor of
$5$ or more larger than those of the neutral resonances. This feature
is less pronounced for the negatively-charged ${\Sigma^*}^- + \gamma^*
\to \Sigma^-$ process. In fig.~\ref{fig:spin_3_2_iso_1_m}, one does
not observe HA's with a magnitude larger than $100 \times
10^{-3}$~GeV$^{-1/2}$.  All the computed EM decay widths contained in
table~\ref{tab:PA_sig_sig_3_2} are smaller than $1.0$~MeV. For the
decuplet member ${\Sigma^*}^-(1385)$, the PDG~\cite{PDG2004} reports
an upper value of $0.01$~MeV. The calculated value of $0.0117$~MeV is
slightly larger. This small discrepancy may be attributed to the fact
that we use the theoretical masses to compute the radiative decay
widths.

\begin{table}
\begin{center}
\caption{Calculated masses, photo-amplitudes and EM decay widths for
  the $\Sigma^* + \gamma \rightarrow \Sigma(1193)$ transitions for
  $J=3/2$ $\Sigma^*$ resonances. The charge of the $\Sigma^*$
  isospin-triplet member is indicated by the superscript
  $^{(0,+,-)}$. Masses and decay widths are given in units of MeV,
  photo-amplitudes are given in units of $10^{-3}$~GeV$^{-1/2}$.}
\label{tab:PA_sig_sig_3_2}
\vspace*{10pt}
\begin{tabular}{|c|c|c|c|c|c|}
\hline\noalign{\smallskip}
Resonance & $M_{calc}$ & $|A_{1/2}|$ & $|A_{3/2}|$ & $\Gamma_{calc}$ \\
\noalign{\smallskip}\hline\hline\noalign{\smallskip}
$P^0_{13}(1385)$ & $1409$ & $27.8$ & $48.0$ & $0.181$ \\
$P^0_{13}(1840)$ & $1902$ & $15.4$ & $-5.25$ & $0.0960$ \\
$P^0_{13}(2080)$ & $1950$ & $14.3$ & $23.7$ & $0.303$ \\
$D^0_{13}(1580)$ & $1675$ & $2.82$ & $32.9$ & $0.230$ \\
$D^0_{13}(1670)$ & $1727$ & $6.77$ & $6.45$ & $0.0214$ \\
$D^{(3) 0}_{13}$ & $1780$ & $-24.4$ & $-25.5$ & $0.349$ \\
\noalign{\smallskip}\hline\noalign{\smallskip}
$P^+_{13}(1385)$ & $1409$ & $62.6$ & $108.2$ & $0.920$ \\
$P^+_{13}(1840)$ & $1902$ & $80.0$ & $-25.6$ & $2.559$ \\
$P^+_{13}(2080)$ & $1950$ & $29.7$ & $48.5$ & $1.280$ \\
$D^+_{13}(1580)$ & $1675$ & $-40.0$ & $65.2$ & $1.235$ \\
$D^+_{13}(1670)$ & $1727$ & $13.1$ & $40.3$ & $0.440$ \\
$D^{(3) +}_{13}$ & $1780$ & $-59.8$ & $-40.8$ & $1.468$ \\
\noalign{\smallskip}\hline\noalign{\smallskip}
$P^-_{13}(1385)$ & $1409$ & $-7.06$ & $-12.2$ & $0.0117$ \\
 & & & & ($< 0.01$) \\
$P^-_{13}(1840)$ & $1902$ & $-47.1$ & $15.1$ & $0.887$ \\
$P^-_{13}(2080)$ & $1950$ & $-1.20$ & $-1.05$ & $0.00101$ \\
$D^-_{13}(1580)$ & $1675$ & $45.7$ & $0.588$ & $0.441$ \\
$D^-_{13}(1670)$ & $1727$ & $0.397$ & $-27.4$ & $0.184$ \\
$D^{(3) -}_{13}$ & $1780$ & $10.9$ & $-10.3$ & $0.0630$ \\
\noalign{\smallskip}\hline
\end{tabular}
\end{center}
\end{table}

\section{Helicity asymmetries}\label{sec:hel_asym}

\begin{figure}
\begin{center}
\resizebox{0.45\textwidth}{!}{\includegraphics{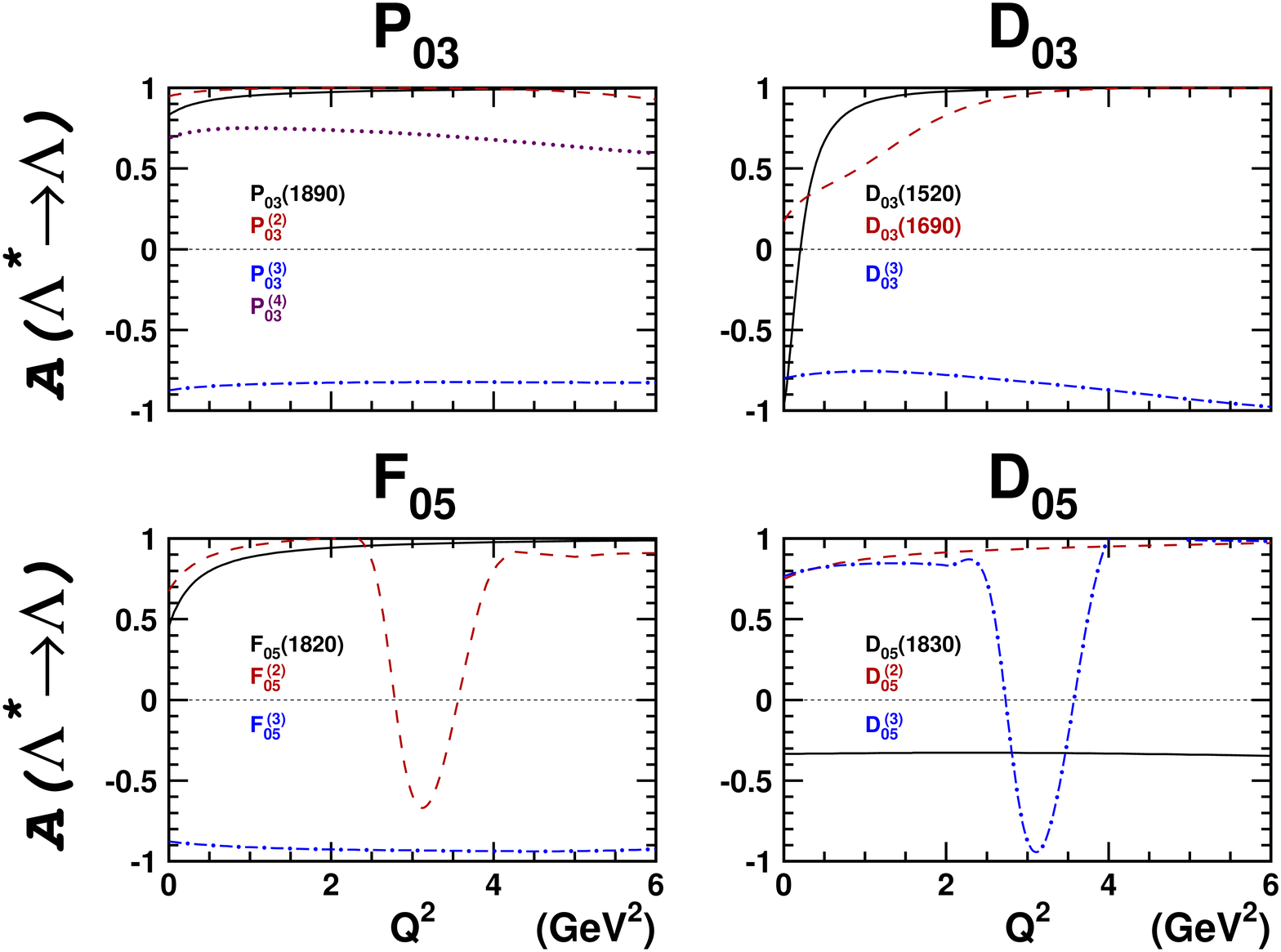}}
\caption{The helicity asymmetry as defined in
  eq.~(\ref{eq:def_ha_asym}) for the lowest-lying spin $J=3/2$ and
  $J=5/2$ $\Lambda$-resonances decaying to the $\Lambda$ ground
  state.}
\label{fig:ha_asym}
\end{center}
\end{figure}

\begin{figure}
\begin{center}
\resizebox{0.45\textwidth}{!}{\includegraphics{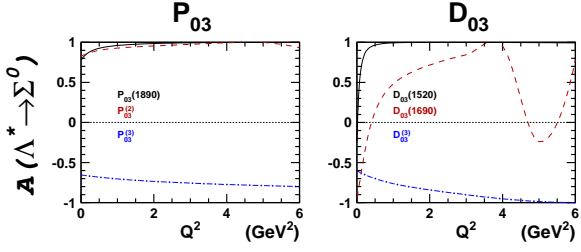}}
\caption{The helicity asymmetry as defined in
  eq.~(\ref{eq:def_ha_asym}) for the lowest-lying spin $J=3/2$
  resonances decaying to the $\Sigma^0$ ground state.}
\label{fig:ha_asym_ext}
\end{center}
\end{figure}

For hyperon resonances with $J \geq 3/2$, the behaviour of the
helicity asymmetries can be qualitatively understood. These
asymmetries are defined analogous to the isospin asymmetries of
eq.~(\ref{eq:asym_sig_lam_lam*})
\begin{equation}
\mathcal{A} = \frac {|A_{1/2}|^2 - |A_{3/2}|^2} {|A_{1/2}|^2 +
  |A_{3/2}|^2} \; .
\label{eq:def_ha_asym}
\end{equation}
The helicity asymmetries of the lowest-lying $J=3/2$ and $J=5/2$
$\Lambda^*$ resonances for the decay to the $\Lambda$ ground state are
shown in fig.~\ref{fig:ha_asym} and the asymmetries for the $J=3/2$
$\Lambda^*$ resonance for the decay to the $\Sigma^0$ ground state are
displayed in fig.~\ref{fig:ha_asym_ext}. In most cases, the helicity
asymmetries approach $+1$ for high $Q^2$, yet for some resonances, the
helicity asymmetry is negative.

To understand the sign of the asymmetries, one can project the
corresponding BS amplitude on the $SU(6)$ spin-flavour basis
states~\cite{greinersymmetries}. This was done in
ref.~\cite{loeringphd}. It turns out that the BS amplitudes of the
resonances for which $\mathcal{A}$ approaches $+1$, receive their
largest contribution from $SU(6)$ spin-flavour states for which the
total spin $S = 1/2$.  On the other hand, the BS amplitudes of the
resonances for which the helicity asymmetry becomes negative at high
$Q^2$, are dominated by $S=3/2$ $SU(6)$ states.

\begin{figure}
\begin{center}
\resizebox{0.45\textwidth}{!}{\includegraphics{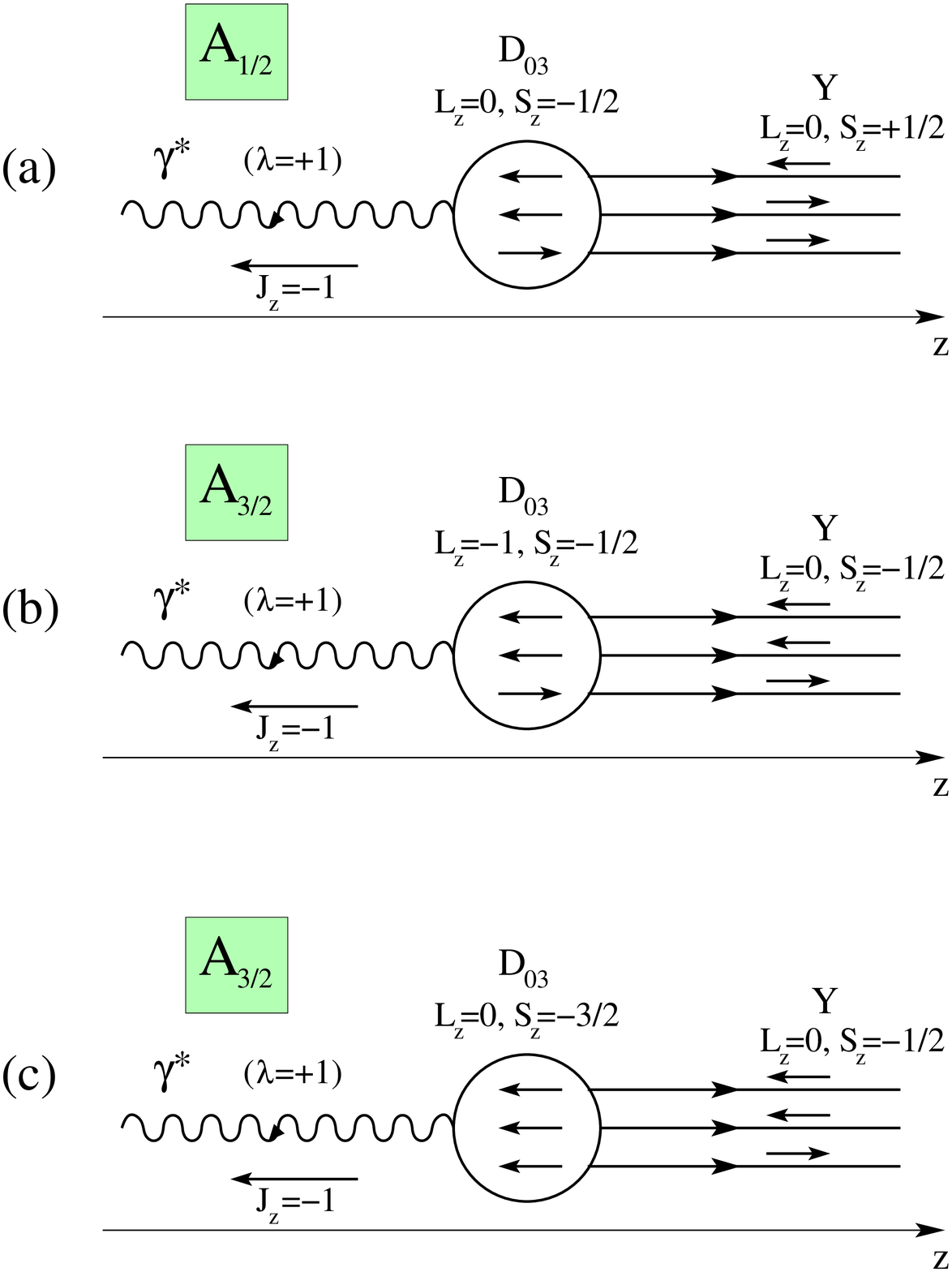}}
\caption[Contributions to the helicity amplitudes for different spin
    states of the hyperon resonance.]{A $D_{03}$ resonance in
    its rest frame decays electromagnetically to a ground-state
    hyperon $Y$. The three depicted processes refer to different
    contributions to the helicity amplitudes $A_{1/2}$ (process (a))
    and $A_{3/2}$ (processes (b) and (c)).}
\label{fig:hel_asym_all}
\end{center}
\end{figure}

This observation can be explained qualitatively by considering the EM
decay of \textit{e.g.} a $D_{03}$ resonance to a ground-state hyperon
$Y$ in the resonance rest frame (see fig.~\ref{fig:hel_asym_all}). For
high $Q^2$, the photon preferentially couples to the individual CQ's,
which means that the major contribution to the $A_{1/2}$ comes from
the process in fig.~\ref{fig:hel_asym_all}(a). There, one of the CQ's
with negative spin projection along the $z$-axis emits a photon of
positive helicity, and flips its spin. This process is allowed for all
$D_{03}$ resonances. When the BS amplitude has its main contributions
from $SU(6)$ states for which $S=1/2$, the major contribution to the
$A_{3/2}$ comes from the process in
fig.~\ref{fig:hel_asym_all}(b). There, one could argue that the photon
is emitted by the resonance as a whole, because the spin projections
of the three CQ's remain unaltered and the projection of the orbital
angular momentum ($L_z$) changes by $1$. As mentioned before, at high
$Q^2$, the photon preferentially couples to the individual quarks. As
a consequence, a process like the one in
fig.~\ref{fig:hel_asym_all}(b) is suppressed relative to the one of
fig.~\ref{fig:hel_asym_all}(a). If the BS amplitude of the resonance
is dominated by $S=3/2$ $SU(6)$ states, at high $Q^2$ the major
contribution to the $A_{3/2}$ comes from the process in
fig.~\ref{fig:hel_asym_all}(c). Here, the photon is emitted by a
single CQ, which accordingly flips its spin. In the situation of
fig.~\ref{fig:hel_asym_all}(c), three CQ's can emit the photon, while
in fig.~\ref{fig:hel_asym_all}(a), only two can do that. Therefore,
the $A_{3/2}$ can be anticipated to be larger than the $A_{1/2}$,
resulting in negative helicity asymmetries.

\begin{figure}
\begin{center}
\resizebox{0.45\textwidth}{!}{\includegraphics{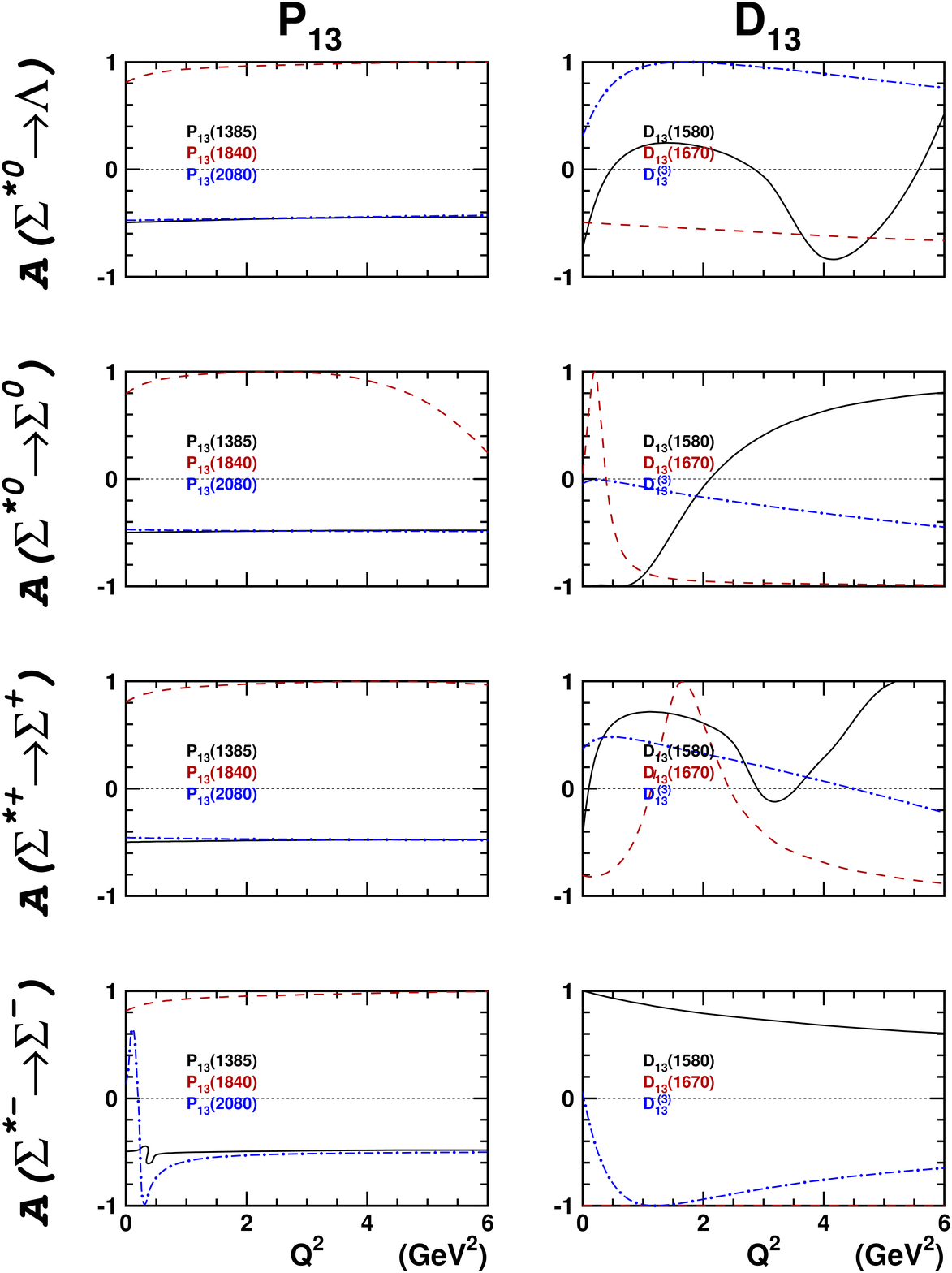}}
\caption{The helicity asymmetry as defined in
  eq.~(\ref{eq:def_ha_asym}) for the lowest-lying spin $J=3/2$
  $\Sigma$-resonances decaying to the $\Lambda$ (upper row), the
  $\Sigma^0$ (second row), the $\Sigma^+$ (third row) and the
  $\Sigma^-$ (lower row) ground states.}
\label{fig:ha_asym_sig}
\end{center}
\end{figure}

A stringent test of the abovementioned argument is provided by the
helicity asymmetries of $J=3/2$ $\Sigma$ resonances, decaying to the
$\Lambda$ and $\Sigma$ ground states. This is illustrated in
fig.~\ref{fig:ha_asym_sig}. The $P_{13}(1385)$, a member of the baryon
decuplet, possesses a symmetric spin wave function. In
ref.~\cite{loering3} it was pointed out that the $P_{13}(2080)$
resonance has an almost purely-symmetric spin wave function. Both
resonances display negative helicity asymmetries, even at relatively
low values of $Q^2$, for all isospin channels. Furthermore, since only
two CQ's contribute to the process in fig.~\ref{fig:hel_asym_all}(a)
and three CQ's contribute to the process of
fig.~\ref{fig:hel_asym_all}(c), one may expect a helicity asymmetry of
$\frac{2^2-3^2}{2^2+3^2} \approx -0.4$. This is clearly in agreement
with the left panels of fig.~\ref{fig:ha_asym_sig}. The computed
helicity asymmetry of the $P_{13}(1840)$ is in accordance with a
purely mixed-symmetric spin wave function reported in
ref.~\cite{loering3}.

For the $D_{13}$ resonances, the situation is more complicated. The
spin wave function of the $D_{13}(1580)$ is dominantly of mixed
symmetry ($S=1/2$), resulting in a helicity asymmetry which goes to
$+1$ at high $Q^2$. The $D_{13}(1670)$ has a spin wave function which
is a mixture of $S=1/2$ and mostly $S=3/2$ components, and thus
displays negative helicity asymmetries. Finally, the $D_{13}^{(3)}$
resonance has a spin wave function which is a mixture of $S=3/2$ and
mostly $S=1/2$ components. Its helicity asymmetries seem to depend on
the isospin and charge of the octet hyperon which it decays to.

\section{Conclusions}\label{sec:conclusions}

The Bonn CQ model has been applied to the computation of helicity
amplitudes of strange baryon resonances. The seven parameters entering
the Bonn model were fitted previously to the masses of the best known
baryons~\cite{loering2,loering3}. Therefore, the present results for the
helicity amplitudes and EM decay widths can be regarded as
predictions. We have calculated the electromagnetic decays $\Lambda^*
\to \Lambda(1116)$, $\Lambda^* \to \Sigma(1193)$, $\Sigma^* \to
\Lambda(1116)$, and $\Sigma^* \to \Sigma(1193)$ for the lowest-lying
$\Lambda^*$'s and $\Sigma^*$'s.

The presented results show interesting features. The first excited
state of a certain spin and parity (and sometimes also higher excited
states) couples con\-si\-dera\-bly stron\-ger to a photon with finite
virtuality $Q^2$ than to a real photon. Therefore, these resonances
can be better studied with virtual photons. Further, the lowest-lying
$\Lambda^*$ seems to decay preferably to the $\Lambda(1116)$, while
the second and third excited $\Lambda^*$ decays preferentially to the
$\Sigma^0(1193)$.

According to the computed helicity amplitudes, the spin $J=5/2$
$F_{05}(1820)$ and $D_{05}(1830)$ $\Lambda$ resonances have a
reasonable EM coupling to the $\Lambda(1116)$. A second resonance with
$J^\pi=5/2^-$, the $D^{(2)}_{05}$ with a computed mass of about
$2100$~MeV, remains unobserved experimentally, but has larger helicity
amplitudes than the first $D_{05}$ resonance. On the basis of these
observations, neglecting $J=5/2$ $\Lambda^*$ resonances in the
$u$-channel background contribution of the \ggsL process should be
done with care.

For the electromagnetic decay of a $\Sigma^*$ resonance to the
$\Sigma$ ground state, three situations, one for each member of the
$\Sigma^*$ isospin triplet, need to be considered. The results show
that the charged states of some $\Sigma^*$ resonances (\textit{e.g.} the
$P_{11}(1660)$, the $S_{11}(1620)$ and the $D_{13}(1580)$) have larger
helicity amplitudes than the neutral state. Therefore, these
$\Sigma^*$ resonances are expected to affect the $p(e,e'K^0)\Sigma^+$
process more than the $p(e,e'K^+)\Sigma^0$ process.

Our investigations lend additional support for the peculiar structure
of the $S_{01}(1405)$, already pointed out in
refs.~\cite{schat95b,jido_garcia-recio,hyodo04}. The predicted EM
decay width is much larger than what is experimentally measured, both
for decay to the $\Lambda(1116)$ and to the $\Sigma(1193)$. In this
respect, we would like to note that the lowest-lying $\Sigma^*$ with
negative parity, the $S_{11}(1620)$, also has large EM decay widths to
the $\Lambda(1116)$ and $\Sigma(1193)$. In contrast to the
$S_{01}(1405)$, the mass of the $S_{11}(1620)$ is well reproduced by
the Bonn CQ model. Furthermore, our predictions for the EM decay
widths of the $D_{03}(1520)$ and $P_{13}(1385)$ resonances seem to be
in good qualitative agreement with the PDG values~\cite{PDG2004}.

We find larger-than-average decay widths for the process $S_{01}(1670)
\to \Sigma(1193) + \gamma$. This explains the fact that the $\Kbar^- p
\to \gamma \Sigma$ cross section is roughly a factor of four larger
than the one for the $\Kbar^- p \to \gamma \Lambda$ reaction for kaon
momenta of about $750$~MeV$/c$ (invariant mass of about
$1678$~MeV)~\cite{prakhov01}. Also the $D_{13}(1670)$ can give a
sizeable contribution to this process.

Finally, the behaviour of the helicity asymmetries for $J \geq 3/2$
resonances lends support for an overall picture in which at high
$Q^2$, the photon couples to an individual constituent quark, rather
than the baryon (resonance) as a whole.

\begin{acknowledgement}
This work is supported by the Research Council of Ghent
University. BCM and HRP acknowledge the support of the European
Community-Research Infrastructure activity under the FP6 ``Structuring
the European Research Area'' program (Hadron Physics, contract number
RII3-CT-2004-506078) and the support within the DFG SFB/TR16
``Subnuclear Structure of Matter - Elektromagnetische Anregung
subnuklearer Systeme''.
\end{acknowledgement}

\appendix

\section{Effective interactions and interaction kernels}\label{sec:potentials}

In this Appendix, a brief description of the quark-quark interactions
used in the kernel of the Bethe-Salpeter equation~(\ref{eq:BSE}) is
given. As shown in sect.~\ref{sec:SE}, the Bethe-Salpeter equation can
be reduced to a Salpeter equation if the interactions are assumed to
be instantaneous. In the model, two types of interactions appear. The
three-particle-irreducible confinement potential will be discussed in
sect.~\ref{sec:conf}. The instanton-induced two-particle-irreducible
residual interaction is the subject of sect.~\ref{sec:thooft}.

\subsection{Confinement potential}\label{sec:conf}

In the Bonn model, the confinement interaction is a string-like
potential which rises linearly with the interquark distances. This
results in almost linear Regge trajectories for both mesons and
baryons in the Bonn model.

The confinement potential is the only three-particle-irreducible
interaction that enters the model. It is given by~:
\begin{multline}
V^{(3)} \left( x_1,x_2,x_3;x'_1,x'_2,x'_3 \right) \ = \
V^{(3)}_{\text{conf}} \left( \mathbf{x}_1,\mathbf{x}_2,\mathbf{x}_3
\right) \\ \times \, \delta^{(1)} \left( x^0_1-x^0_2 \right) \
\delta^{(1)} \left( x^0_2-x^0_3 \right) \ \delta^{(4)} \left( x_1-x'_1
\right) \\ \times \, \delta^{(4)} \left( x_2-x'_2 \right) \
\delta^{(4)} \left( x_3-x'_3 \right) \; . \;
\label{eq:3p_irr_conf}
\end{multline}
Here, the one-dimensional $\delta$-functions of the time-components
implements the assumption that the interaction is instantaneous. The
actual confinement potential
$V^{(3)}_{\text{conf}}(\mathbf{x}_1,\mathbf{x}_2,\mathbf{x}_3)$ is a
function of the relative quark coordinates, but also comprises
\emph{Dirac} structures which act on the quark spinors. It can be
written as~\cite{loering2}~:
\begin{equation}
V^{(3)}_{\text{conf}} \; = \; a \ \mW_{\text{off}} \; + \; b \
r_{3q}(\mathbf{x}_1,\mathbf{x}_2,\mathbf{x}_3) \ \mW_{\text{str}} \; , \;
\label{eq:Vconf_par_dir_struc}
\end{equation}
where $a$ and $b$ are the confinement parameters, $r_{3q}$ is a
measure for the interquark distance, and $\mW_{\text{off}}$ and
$\mW_{\text{str}}$ are the Dirac structures operating on the
constituent-quark spinors. The parameters $a$ and $b$ are the sole
parameters associated with the confinement potential. These parameters
and the $m_u \equiv m_d \equiv m_n$ nonstrange constituent-quark mass
are determined by optimizing the model results for the $\Delta$
spectrum to the experimentally best-known resonance masses. The
optimized $a$, $b$ and $m_n$ are contained in
table~\ref{tab:par_bonn}.

\begin{table}
\begin{center}
\caption{The seven parameters of the Bonn model are the
  constituent-quark masses, the confinement offset and slope, the
  't~Hooft interaction range, and the 't~Hooft nonstrange-nonstrange
  and nonstrange-strange interaction strength.}
\label{tab:par_bonn}
\begin{tabular}{||c|c|r|l||}
\hhline{|t:====:t|}
Parameter & Symbol & Value & Unit \\
\hhline{|:=:=:=:=:|}
nonstrange CQ mass & $m_n$ & $330$ & MeV \\
strange CQ mass & $m_s$ & $670$ & MeV \\
\hhline{||-|-|-|-||}
confinement offset & $a$ & $-744$ & MeV \\
confinement slope & $b$ & $470$ & MeV fm$^{-1}$ \\
\hhline{||-|-|-|-||}
't~Hooft $nn$ strength & $g_{nn}$ & $136$ & MeV fm$^3$ \\
't~Hooft $ns$ strength & $g_{ns}$ & $94$ & MeV fm$^3$ \\
't~Hooft range & $\Lambda$ & $0.40$ & fm \\
\hhline{|b:====:b|}
\end{tabular}
\end{center}
\end{table}

\begin{figure*}
\begin{center}
\mbox{\subfigure[]{\resizebox{0.25\textwidth}{!}{\includegraphics{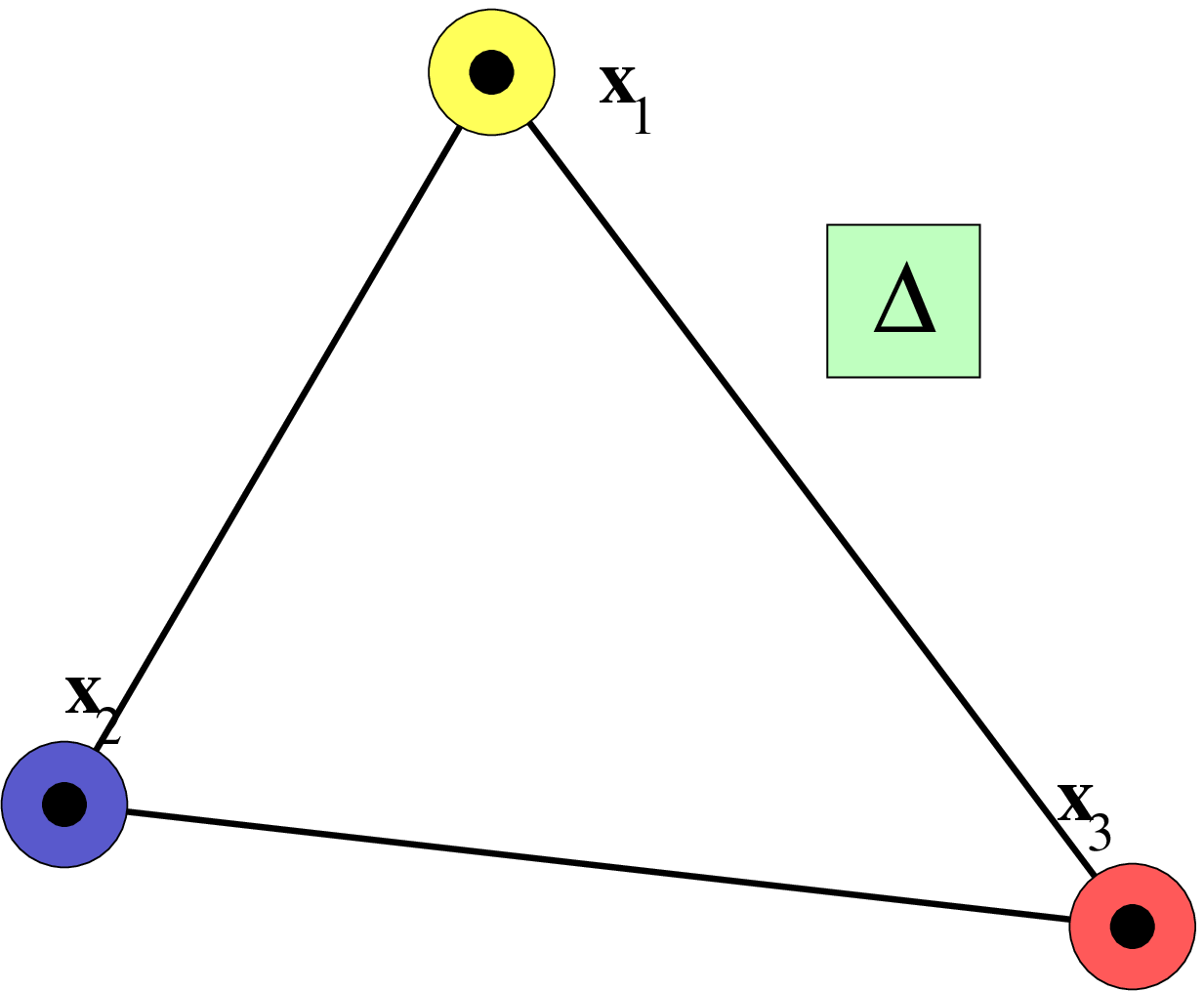}}}\qquad
\subfigure[]{\resizebox{0.25\textwidth}{!}{\includegraphics{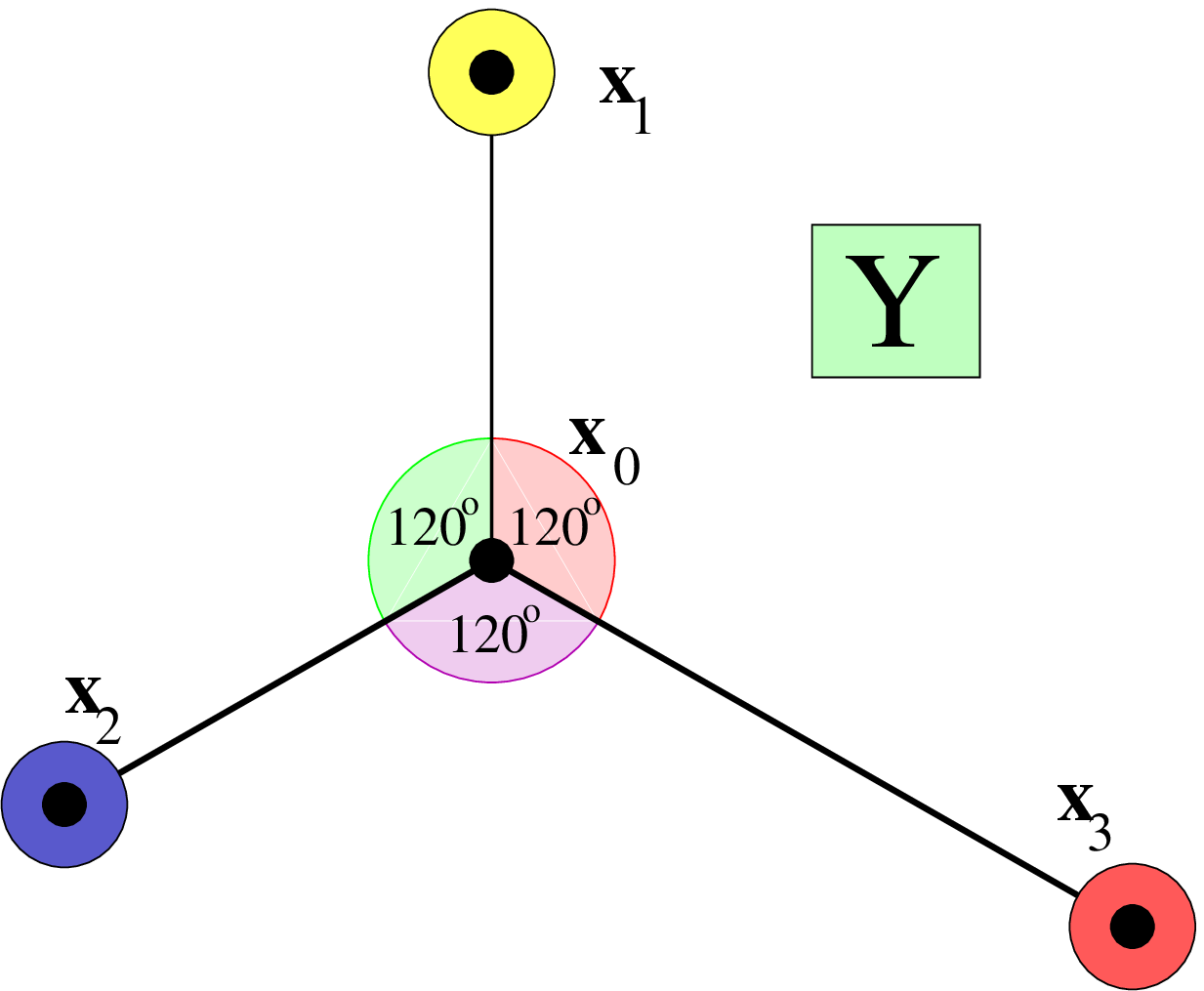}}}\qquad
\subfigure[]{\resizebox{0.25\textwidth}{!}{\includegraphics{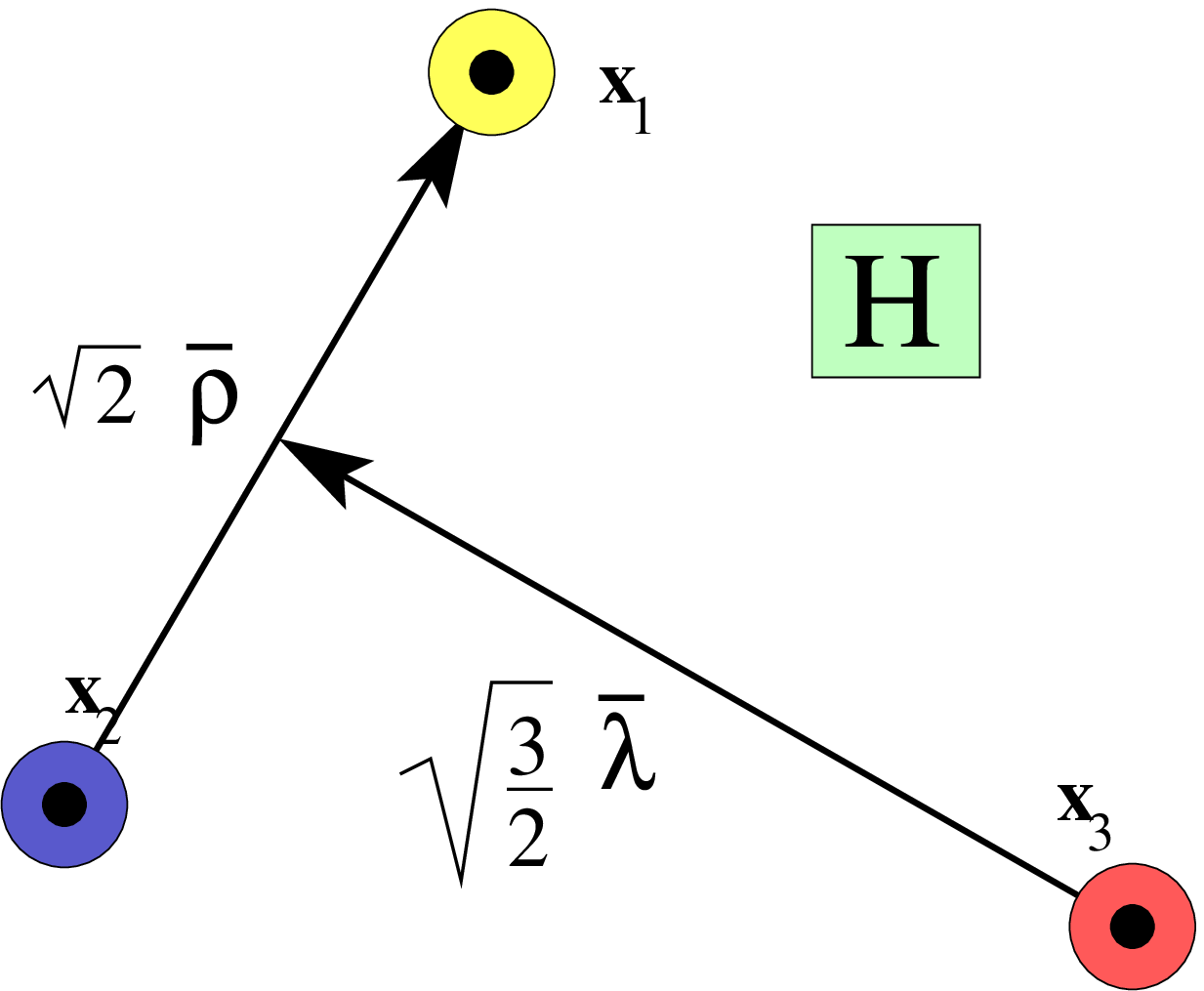}}}}
\caption{The $\Delta$-, the $Y$- and the $H$-configuration for
    the interquark distance $r_{3q}$ are shown in (a), (b) and (c)
    respectively. In (a) and (b), the length of the connecting lines
    is the interquark distance. In (c), $r_{3q}$ is given by
    eq.~(\ref{eq:r3q_def_H}).}
\label{fig:conf_conf}
\end{center}
\end{figure*}

The interquark distance $r_{3q}$ for three constituent quarks
can be defined in different manners. We use the sum of the three
distances between the quarks, which is commonly referred to as a
$\Delta$-configuration~:
\begin{equation}
r_{3q} \left( \mathbf{x}_1,\mathbf{x}_2,\mathbf{x}_3 \right) \ = \
\sum_{i<j} \ |\mathbf{x}_i - \mathbf{x}_j| \; . \;
\label{eq:r3q_def_D}
\end{equation}
In literature, one finds alternative definitions, such as the $Y$- and
$H$-configuration as depicted in fig.~\ref{fig:conf_conf}. The
$Y$-configuration uses the minimal length to connect three points~:
\begin{equation}
r_{3q} \left( \mathbf{x}_1,\mathbf{x}_2,\mathbf{x}_3 \right) \ = \
\min_{\mathbf{x}_0} \ \sum_{i<j} \ |\mathbf{x}_i - \mathbf{x}_0| \; ,
\; \label{eq:r3q_def_Y}
\end{equation}
whereas the $H$-version puts forward the hyperradius as a measure of
the interquark distance~:
\begin{equation}
r_{3q} \left( \mathbf{x}_1,\mathbf{x}_2,\mathbf{x}_3 \right) \ = \
\sqrt{|\mathbf{\rho}|^2 + |\mathbf{\lambda}|^2} \; , \;
\label{eq:r3q_def_H}
\end{equation}
where $\mathbf{\rho} = \frac{1}{\sqrt{2}} (\mathbf{x}_1 -
\mathbf{x}_2)$ and $\mathbf{\lambda} = \frac{1}{\sqrt{6}}
(\mathbf{x}_1 + \mathbf{x}_2 - 2\mathbf{x}_3)$. It turns out, however,
that the slope parameter $b$ of the confinement potential can be
scaled such that the results for the three variants are of equal
quality. Lattice calculations seem to favor a configuration which is a
mixture of the $\Delta$- and $Y$-variant~\cite{lattice}. Numerically,
the $\Delta$-configuration is easier to handle in CQ-model
calculations and is the one adopted here.

The Dirac structures $\mW_{\text{off}}$ and $\mW_{\text{str}}$ have a
large impact on the computed baryon spectrum. The specific choice for
these structures is constrained by the observation that the spin-orbit
effects in the baryon spectrum are moderate (note \textit{e.g.} the
small mass difference between the $S_{11}(1535)$ and $D_{13}(1520)$
nucleon resonances). Furthermore, $\mW_{\text{off}}$ and
$\mW_{\text{str}}$ can have a different structure. The best choice
is~\cite{loeringphd}~:
\begin{subequations}
\begin{multline}
\mW_{\text{off}} \, = \, \frac{3}{4} \ \bigl[ \ \Id \otimes \Id \otimes \Id \ +
  \ \gamma^0 \otimes \gamma^0 \otimes \Id \phantom{\bigr]} \\
  \phantom{\bigl[} + \textrm{cycl. perm.} \ \bigr] \; , \;
  \label{eq:dirac_struc_A_off}
\end{multline}
\begin{multline}
\mW_{\text{str}} \, = \, \frac{1}{2} \ \bigl[ \ - \Id \otimes \Id
  \otimes \Id \ + \ \gamma^0 \otimes \gamma^0 \otimes \Id \ \phantom{\bigr]}
  \\ \phantom{\bigl[} + \textrm{cycl. perm.} \ \bigr] \; . \;
  \label{eq:dirac_struc_A_str}
\end{multline}
\label{eq:dir_struc_A}
\end{subequations}
With this specific choice for the Dirac structures, the
$V^{(3)}_{\text{conf}}$ of eq.~(\ref{eq:Vconf_par_dir_struc}) reduces
to a spin-independent linear confinement potential in the
nonrelativistic limit~\cite{loeringphd}.

\subsection{'t Hooft instanton induced interaction}\label{sec:thooft}

In the Bonn model, the hyperfine splittings in the baryon spectrum are
induced by a two-particle irreducible interaction based on the effects
of instantons on the propagation of light quarks. Instantons are
\emph{classical}, non-perturbative solutions of the QCD Yang-Mills
equations in Euclidean spacetime. They are localized in space and
imaginary time and describe tunneling events. Instantons
(anti-instantons) absorb right-handed (left-handed) light-flavoured
quarks, and emit left-handed (right-handed) ones. As such they mediate
a force between light quarks. Furthermore, instantons change the axial
charge of the QCD vacuum in the presence of an external fermion
source. Therefore, they provide an explanation for the
non-conservation of axial charge. The crucial properties of instantons
were discovered by 't~Hooft~\cite{thooft76}. Therefore, the resulting
interaction between light quarks is sometimes referred to as the
't~Hooft interaction.

The two-body part of the 't~Hooft instanton-induced interaction,
$V^{(2)}_{\text{III}}$, induces a flavour-, spin- and
colour-de\-pen\-dent force between two light quarks. In particular its
acts between flavour antisymmetric quark pairs according to
\begin{multline}
V^{(2)}_{\text{III}} \left( x_1,x_2;x'_1,x'_2 \right) =
V^{(2)}_{\text{'t~Hooft}} \left( \mathbf{x}_1 - \mathbf{x}_2 \right)
\\ \times \, \delta^{(1)} (x^0_1 - x^0_2) \delta^{(4)} (x_1 - x'_1)
\delta^{(4)} (x_2 - x'_2) \; .  
\label{eq:V_residual}
\end{multline}
The 't Hooft two-body potential, $V^{(2)}_{\text{'t~Hooft}}$, is a
function of the distance between the two constituent quarks
$(\mathbf{x}_1-\mathbf{x}_2)$, and comprises the appropriate Dirac
structure and projectors in Dirac- ($\mD$), flavour- ($\mF$) and
colourspace ($\mC$)~:
\begin{multline}
V^{(2)}_{\text{'t~Hooft}} \left( \mathbf{x}_1 - \mathbf{x}_2 \right) =
- 4 \ v_{reg} \left( \mathbf{x}_1 - \mathbf{x}_2 \right) \\ \times \;
\mathcal{P}^{\mathcal{D}}_{S_{12}=0} \ \otimes \ \left( g_{nn}
\mathcal{P}^{\mathcal{F}}_{\mathcal{A}} (nn) + g_{ns}
\mathcal{P}^{\mathcal{F}}_{\mathcal{A}} (ns) \right) \ \otimes \
\mathcal{P}^{\mathcal{C}}_{\bar{3}} \\ \times \, \left( \Id \otimes
\Id + \gamma^5 \otimes \gamma^5 \right) \; .  
\label{eq:V_tHooft}
\end{multline}
Here, $v_{reg} \left( \mathbf{x}_1 - \mathbf{x}_2 \right)$ is a
regulating function, describing the three-dimensional extension of the
interaction~:
\begin{equation}
v_{reg} \left( \mathbf{x} \right) \ = \ \frac{1}{\Lambda^3
  \pi^{\frac{3}{2}}} \ e^{-\frac{|\mathbf{x}|^2}{\Lambda^2}} \; . \;
  \label{eq:v_reg_thooft}
\end{equation}
The range of the interaction, $\Lambda$, is a free parameter in the
Bonn model. It is extracted from a fit of the model results to the
best-known nucleon masses, and its value is listed in
table~\ref{tab:par_bonn}. The magnitude of $\Lambda$ corresponds
roughly to the average size of the
instanton~\cite{loering2,diakonov_shuryak}. The two interaction
strengths $g_{nn}$ and $g_{ns}$, associated with the antisymmetric
nonstrange-nonstrange and the strange-nonstrange flavour projectors,
are also fitting parameters. The $g_{nn}$ coupling strength is fitted
to the nucleon spectrum, and reproduces the hyperfine splitting
between the nucleon and $\Delta(1232)$ resonance. In contrast to the
nucleon, the $\Delta(1232)$ has a symmetric-spin wave
function. Therefore, the $V^{(2)}_{\text{III}}$ of
eq.~(\ref{eq:V_residual}) affects only the nucleon, lowering its mass
compared to the $\Delta$ resonance. The strange-nonstrange coupling
($g_{ns}$) and the strange constituent-quark mass ($m_s$) parameters
are determined in order to reproduce the masses of the experimentally
best-known hyperons. As a matter of fact, the $g_{ns}$ coupling is
responsible for the $\Sigma^*_{(dec.)} - \Sigma_{(oct.)}$ and
$\Xi^*_{(dec.)} - \Xi_{(oct.)}$ mass splittings. Values of $g_{nn}$,
$g_{ns}$ and $m_s$ are listed in table~\ref{tab:par_bonn}.

\end{document}